\documentclass{article}

\usepackage{arxiv}

\usepackage[utf8]{inputenc} 
\usepackage[T1]{fontenc}    
\usepackage{hyperref}       
\usepackage{url}            
\usepackage{booktabs}       
\usepackage{amsfonts}       
\usepackage{nicefrac}       
\usepackage{microtype}      
\usepackage{lipsum}		
\usepackage{graphicx}
\usepackage{natbib}
\usepackage{doi}

\usepackage{fancyvrb}
\usepackage{amsthm}
\usepackage{amsmath}
\usepackage{algorithmic}

\theoremstyle{definition}
\newtheorem{definition}{Definition}

\theoremstyle{plain}
\newtheorem{example}{Example}

\newcommand\Tau{\mathcal{T}}

\usepackage{tikz}
\usetikzlibrary{shapes.geometric, arrows}
\usetikzlibrary{shapes}
\usetikzlibrary{positioning} 
\usetikzlibrary{fit}
\usetikzlibrary{shadows} 
\usetikzlibrary{calc} 
\usetikzlibrary{decorations.markings}
\usetikzlibrary{arrows.meta, bending, intersections}

\tikzset{splrect/.style = {rectangle split, rectangle split horizontal,
                           rectangle split parts=5, minimum height=1cm, 
                           align=center, draw=black},
         rect/.style = {rectangle, rounded corners, minimum width=3cm,
                        minimum height=1cm,text centered, text width=3cm,
                        draw=black},
         rect2/.style = {rectangle, rounded corners, minimum width=4cm,
                        minimum height=1cm,text centered, text width=4cm,
                        draw=black},
         redrect/.style = {rectangle, rounded corners, minimum width=3cm,
                           minimum height=1cm,text centered, text width=3cm,
                           draw=black, fill=red, fill opacity=0.5},
         freerect/.style = {rectangle, rounded corners, minimum width=10cm, minimum height=1cm, draw=black, text centered, text width=10cm},
         arrow/.style = {->,shorten >=1pt,>=stealth',semithick},
         labl/.style = {minimum width=3cm,
                        minimum height=1cm,text centered, text width=3cm,
                        draw=black!0},
         doc/.style={draw, minimum height=4em, minimum width=3em, text width=3cm, text centered, fill=white, double copy shadow={shadow xshift=4pt, shadow yshift=4pt, fill=white, draw}},
         decision/.style={diamond, minimum width=3cm, minimum height=1cm, text centered, draw=black}
}

\tikzset{
  invisible/.style={opacity=0},
  visible on/.style={alt={#1{}{invisible}}},
  alt/.code args={<#1>#2#3}{%
    \alt<#1>{\pgfkeysalso{#2}}{\pgfkeysalso{#3}} 
  },
}

\title{A Benchmark Generator for Combinatorial Testing}

\date{Submitted to IEEE Transactions on Software Engineering on September 21, 2022}

\author{ \href{https://orcid.org/0000-0001-7727-2766}{\includegraphics[scale=0.06]{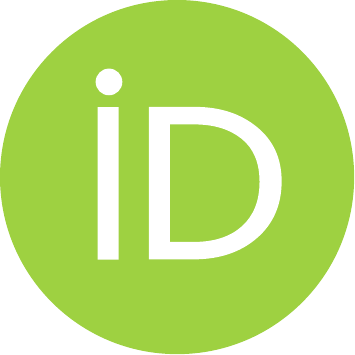}\hspace{1mm}Carlos Ansotegui} \\
    Logic \& Optimization Group \\
	Universitat de Lleida\\
	\texttt{carlos.ansotegui@udl.cat} \\
	\And
	\href{https://orcid.org/0000-0002-3136-7513}{\includegraphics[scale=0.06]{orcid.pdf}\hspace{1mm}Eduard Torres} \\
    Logic \& Optimization Group \\
	Universitat de Lleida\\
	\texttt{eduard.torres@udl.cat} \\
}

\renewcommand{\shorttitle}{A Benchmark Generator for Combinatorial Testing}

\hypersetup{
pdftitle={A Benchmark Generator for Combinatorial Testing},
pdfsubject={},
pdfauthor={Carlos Ansotegui, Eduard Torres},
pdfkeywords={Combinatorial Testing, Benchmark Generator, Covering Arrays, Satisfiability},
}

\begin{document}
\maketitle

\begin{abstract}
	Combinatorial Testing (CT) tools are essential to test properly a wide range of systems (train systems, Graphical User Interfaces (GUIs), autonomous driving systems, etc). While there is an active research community working on developing CT tools, paradoxically little attention has been paid to making available enough resources to test the CT tools themselves. In particular, the set of available benchmarks to asses their correctness, effectiveness and efficiency is rather limited. In this paper, we introduce a new generator of CT benchmarks that essentially borrows the structure contained in the plethora of available Combinatorial Problems from other research communities in order to create meaningful benchmarks. We additionally perform an extensive evaluation of CT tools with these new benchmarks. Thanks to this study we provide some insights on under which circumstances a particular CT tool should be used.
\end{abstract}

\keywords{Combinatorial Testing \and Benchmark Generator \and Covering Arrays \and Satisfiability}

\section{Introduction}

Technology has become an essential part of our daily lives. Unfortunately, bugs, errors and faults are also highly tied to technology systems.  Therefore, one of the main goals of any system's designer is to efficiently and effectively test these systems to detect and fix as many of these faults in a controlled environment prior to the deployment of the real application.

More precisely, by System Under Test (SUT), we refer to a black box with a set of input parameters $P$, which take values from a finite domain, and performs some task that we can check to confirm the SUT is working as designed. To model that there are configurations of values for these input parameters which are \emph{not allowed} we can use a set of SUT constraints $\varphi$ that encode these \emph{forbidden} configurations.

The term Combinatorial Testing (CT)~\cite{nie2011survey} encompasses testing techniques in which multiple combinations of the input parameters are used to perform testing of the software product. The main idea behind CT approaches is that, in general, errors are produced by the interaction of a \emph{relatively small} set of parameters~\cite{KuhnWG04}. Therefore, a reasonable number of tests can suffice to detect bugs in the SUT.

Paradoxically, the main drawback when designing CT tools, which are used to test other tools, is the absence of a rich collection of SUT benchmarks to test empirically the correctness and goodness of these CT tools. This is particularly true when we deal with CT tools for SUTs with constraints since unless we obtain benchmarks (SUTs with constraints) from the industry field, we have to \emph{artificially} generate them. Moreover, we also want to control the hardness of the constraints associated with the SUT.

In the literature, we find a limited collection of benchmarks for SUT with constraints.
In~\cite{cohen2008constructing} we find 5 real-world benchmarks and 30 artificially generated benchmarks considering the features of the real-world ones.
In~\cite{segall11} the authors provide 20, relatively easy, real-world benchmarks, and in~\cite{Yu15,Yamada16} we find two and one real-world benchmark respectively.
Finally, in~\cite{10.1145/3236024.3236067} there are 116 versions of 48 real-world benchmarks.

The absence of a rich collection of benchmarks, in contrast to other disciplines, such as in the Satisfiability research community \cite{DBLP:series/faia/336}, is a source of an endless list of disadvantages. This prevents the CT community from having a deeper understanding of the competitiveness of CT tools and slows down the development of new algorithms and tools. Moreover, current CT tools can be biased toward solving problems that are similar to the reduced set of available benchmarks, and, last but not least, makes it difficult to run competitions in the field, events that historically have boosted the research in other disciplines\footnote{The CT community has recently organized also a competition: \url{https://fmselab.github.io/ct-competition/Competition2022.html}}.

In the SAT community, the automatic generation of SAT instances close to real-world problems is actually listed in~\cite{10.5555/1624162.1624170} as CHALLENGE 10: "Develop a generator for problem instances that have computational properties that are more similar to real-world instances". It comes as no surprise we have a similar challenge in the CT community.

In this paper, we present a new generator for SUTs with constraints. We take an original approach since the SUT will actually be a \emph{subproblem} of an existing decision combinatorial problem. In particular, we will focus on the SAT problem and SAT instances. In this sense, we fix some variables in the SAT instance and simplify the formula by applying some incomplete inference Boolean mechanisms such as Unit Propagation. From the variables in the new remaining subformula, we select some Boolean variables to be the input parameters of the new SUT and the rest to be auxiliary variables.  The clauses in the subformula become the constraints of the new SUT.

Our approach allows in a natural way access to the rich diversity of constraints that belong to the plethora of industrial, crafted and random SAT instances available in the SAT community. 

Moreover, the current SUT benchmarks available in the literature do contain a set of constraints that we can consider \emph{easy} from the point of view of its computational hardness. We should expect these sets of constraints to be \emph{harder} as CT techniques are applied to more complex real-world scenarios. In this sense, we show how to control empirically the hardness of the constraints included in the SUT.

Equipped with this new generator of SUTs with constraints, we generate a new set of SUT benchmarks and conduct an extensive evaluation on some of the available CT tools. In particular, we focus on the evaluation of the IPOG~\cite{https://doi.org/10.1002/stvr.381} and BOT-its~\cite{DBLP:conf/cp/AnsoteguiOT21,Yamada16} algorithms.
Thanks to our experimental investigation we are able to come up with some new interesting recipes on when to use a particular CT tool to test a given SUT.
As an additional contribution, we will make available to the community the generated benchmarks as well as the generator.

This paper is structured as follows:
Section~\ref{sec:gen-bench-sat} presents our approach to generate SUT with constraints benchmarks.
First of all, we start by introducing several definitions and examples related to CT.
Then, we review some of the most widely-used formats to represent SUT with constraints and extend one of them in Section~\ref{sec:new-sut-format} to make more convenient the representation of the generated SUT constraints.
Then, in Section~\ref{sec:sut-sat-gen} we describe in detail the \emph{SUT-G} generator.
In Section~\ref{sec:algs} we summarize two well-known algorithms for building Mixed Covering Arrays with Constraints (MCACs), a particular CT problem.
Later, in our experimental evaluation (Section~\ref{sec:exps-ipog-bot}), we use these algorithms to solve benchmarks generated with the \emph{SUT-G} generator to try to provide another point of view on the behaviour of these MCAC algorithms.
Thanks to this experimental evaluation, we are able to provide in Section~\ref{sec:recipes-mcac} several recipes to effectively apply MCAC generation tools to real-world scenarios.
Finally, we conclude this work in Section~\ref{sec:conclusions}.


\section{Generating SUTs with Constraints}
\label{sec:gen-bench-sat}

In this section, we present the \emph{SUT-G} generator, a novel benchmark generator that can craft Systems Under Test (SUT) models with constraints from any SAT instance.
First of all, we review some of the current state-of-the-art formats to represent SUTs with constraints and propose an extended format to represent SUT constraints more conveniently in some scenarios.
Finally, we describe the \emph{SUT-G} generator, the second main contribution of this paper.

To start, we introduce some definitions and examples.

\begin{definition}
A System Under Test (SUT) model is a tuple $\langle P,\varphi \rangle$, where 
$P$ is a finite set of variables $p$ of finite domain, called SUT parameters, and $\varphi$ is a set of constraints on $P$, called SUT constraints, that implicitly represents the parameterizations that the system accepts. 
We denote  by $d(p)$ and $g_p$, respectively, the domain and the cardinality domain of $p$.
For the sake of clarity, we will assume that the system accepts at least one parameterization.    
\end{definition}

In the following, we assume $S=\langle P,\varphi \rangle$ to be a SUT model. We will refer to $P$ as $S_P$, and to $\varphi$ as $S_{\varphi}$.

\begin{example} \label{ex:sut}
Consider an online service that offers a web page and a mobile application that share the same code base. We want to find potential failures given different OS, platforms, screen resolutions and orientations. The definition of this example SUT $S$ regarding its parameters $S_P$ is the following:

\begin{align*}
OS\ (OS) \in \{&Linux\ (L),\ Windows\ (W),\\
&Mac\ (M),\ iOS\ (i),\ Android (A)\} \\
Platform\ (Pl) \in \{&Firefox\ (F),\ Safari\ (S),\\
&Chrome\ (C),\ App\ (A)\} \\
Resolution\ (Re) \in \{&4K\ (K),\ FHD\ (F),\ HD\ (H),\\ 
&WXGA\ (W)\} \\
Orientation\ (Or) \in \{&Portrait\ (P),\ Landscape\ (L)\}
\end{align*}
\end{example}

In this scenario we also have a set of SUT constraints $S_\varphi$:

\begin{equation} \label{eq:sut-constr-1}
    ((OS = L) \lor (OS = W) \lor (OS = M)) \rightarrow ((Or = L) \land (Pl \neq A))
\end{equation}
\begin{equation} \label{eq:sut-constr-2}
    (Pl=S) \rightarrow ((OS=M) \lor (OS=i))
\end{equation}
\begin{equation}\label{eq:sut-constr-3}
    ((OS=i) \lor (OS=A)) \rightarrow (Re \neq K)
\end{equation}

Whenever the OS is \emph{Linux}, \emph{Windows} or \emph{Mac}, the Orientation must be \emph{Landscape} and the Platform cannot be \emph{App} (equation \eqref{eq:sut-constr-1}). When the Platform is \emph{Safari}, the OS must be \emph{Mac} or \emph{iOS} (equation \eqref{eq:sut-constr-2}). Finally, for \emph{iOS} and \emph{Android} the Resolution cannot be \emph{4K} (equation \eqref{eq:sut-constr-3}).

\subsection{Available formats for representing SUTs}
\label{sec:avail-formats-suts}

We review four of the most widely used formats for representing SUT models.

\subsubsection{The CASA format}
\label{sec:casa-format}

CASA~\cite{5033175} is CT tool for building Covering Arrays through Simulated Annealing.
This tool defines a SUT format\footnote{The full CASA format specification can be found in \url{http://cse.unl.edu/~citportal/citportal/academic}} that is still used by some recent tools such as WCA~\cite{FU2020106288} or AutoCCAG~\cite{9402109}.

This format consists of two separate files: the \texttt{.model} file, which defines the SUT parameters and their domains, and the \texttt{.constraints}, which define the SUT constraints.

For the \texttt{.model} file, we must define the number of parameters of the SUT, the desired \emph{strength} of the interactions, and the domain of the parameters.
No more information can be added to the \texttt{.model} file.
In Figure~\ref{fig:casa-model} we show the associated CASA model file for Example~\ref{ex:sut} and strength $t=2$.

\begin{figure}[ht]
\centering
\begin{verbatim}
4
2
5 4 4 2
\end{verbatim}
\caption{CASA model file for Example~\ref{ex:sut}}
\label{fig:casa-model}
\end{figure}

Regarding constraints, these can only be represented as conjunctions of disjunctions (i.e. CNF format).
We must specify the total number of clauses and for each clause its number of symbols.
These symbols (also named as literals) can only refer to parameters' values, and they are derived from the parameter order defined in the model file, starting from 0.
Literals can be positive or negative.
For example, in Example~\ref{ex:sut}, $OS=Linux$ would be symbol 1, $OS=Windows$ symbol 2, and so on.
Figure~\ref{fig:casa-constraints} shows the CASA constraints for Example~\ref{ex:sut}, where we have manually converted the SUT constraints to CNF.

\begin{figure}[ht]
\centering
\begin{verbatim}
9
2
- 0 + 14
2
- 1 + 14
2
- 2 + 14
2
- 0 - 8
2
- 1 - 8
2
- 2 - 8
2
- 4 - 9
2
- 3 - 9
3
- 6 + 2 + 3
\end{verbatim}
\caption{CASA constraints file for Example~\ref{ex:sut}}
\label{fig:casa-constraints}
\end{figure}

Notice that this format for SUT constraints does not allow auxiliary variables (i.e. all the literals that appear in the constraints must be part of the defined parameters).
As we will discuss in Section~\ref{sec:acts-format}, auxiliary variables are very useful to represent certain SUT constraints in a more convenient way.

\subsubsection{The PICT format}
\label{sec:pict-format}

PICT~\cite{Czerwonka06} is another CT tool that also defines its own format for representing SUT models\footnote{The full specification for the PICT format can be found here: \url{https://github.com/microsoft/pict/blob/main/doc/pict.md}}.

This is a more advanced format than CASA (see Section~\ref{sec:casa-format}) and offers more features in terms of the definition of the SUT.

In this case, all the SUT is defined using a single \texttt{.pict} file.
Unlike in CASA, in PICT we can define the name of the parameter, its type (\emph{string} or \emph{numeric}) and, for \emph{strings}, the name of the parameter's values.
This can be useful to better understand the definition of the SUT, as well as the output test suite.

Regarding SUT constraints, PICT offers \emph{if-then} operands, as well as negations, conjunctions and disjunctions.
SUT constraints are the conjunction of all the defined constraints.

PICT also provide additional features for its format such as negative testing or weighting.
Its full definition can be found here: \url{https://github.com/microsoft/pict/blob/main/doc/pict.md}

Figure~\ref{fig:pict-example} shows Example~\ref{ex:sut} expressed in PICT format.

\begin{figure}
\centering
\begin{verbatim}
OS: L, W, M, i, A
Pl: F, S, C, A
Re: K, F, H, W
Or: P, L

IF ([OS] = "L" OR [OS] = "W" OR [OS] = "M")
THEN ([Or] = "L" AND [Pl] <> "A")
IF [Pl] = "S" THEN ([OS] = "M" OR [OS] = "i")
IF ([OS] = "i" OR [OS] = "A") THEN [Re] <> "K"
\end{verbatim}
\caption{Representation of SUT in Example~\ref{ex:sut} in PICT format}
\label{fig:pict-example}
\end{figure}

\subsubsection{The CTWedge format}
\label{sec:ctwedge-format}

CTWedge~\cite{8411769} is a CT tool that also defines its own SUT format\footnote{The grammar of the CTWedge format can be found here: \url{https://github.com/fmselab/ctwedge/blob/master/ctwedge.parent/ctwedge/src/ctwedge/CTWEdge.xtext}}.

As in PICT (see Section~\ref{sec:pict-format}), CTWedge also allows to name parameters and values, as well as to define their type (which can be \emph{Bool}, \emph{Enumerative} and \emph{Range}).

To define constraints, it also allows to use negation, conjunction, disjunction and implication operands, and SUT constraints are the conjunction of all the defined constraints.
As in all the previous cases, this format also does not support auxiliary variables.

Figure~\ref{fig:ctwedge-example} shows Example~\ref{ex:sut} represented using the CTWedge format.

\begin{figure}
\centering
\begin{verbatim}
Model MySUT

OS: { L W M i A }
Pl: { F S C A }
Re: { K F H W }
Or: { P L }

Constraints:
    # (OS == "L" || OS == "W" || OS == "M") =>
        (Or == "L" && Pl != "A") #
    # Pl == "S" => (OS == "M" || OS== "i") #
    # (OS == "i" || OS == "A") => Re != "K" #
\end{verbatim}
\caption{Representation of SUT in Example~\ref{ex:sut} in CTWedge format}
\label{fig:ctwedge-example}
\end{figure}

\subsubsection{The ACTS format}
\label{sec:acts-format}

ACTS~\cite{BorazjanyYLKK12} is one of the most widely used CT tools for building Covering Arrays.
It implements several CT algorithms, and also defines its own format for defining SUTs\footnote{The full specification of the ACTS format can be found in \url{https://csrc.nist.rip/groups/SNS/acts/documents/acts_user_guide_2.92.pdf}}.

As in PICT and CTWedge (Sections~\ref{sec:pict-format} and~\ref{sec:ctwedge-format}), in ACTS we can define the name of the parameter, its type (int, enum or bool) and, for \emph{enums}, the name of the parameter's values.

Regarding SUT constraints, ACTS allows the usage of implications, conjunctions and disjunctions to define a constraint.
SUT constraints are defined as the conjunction of each of the defined constraints.
All the symbols that appear in the SUT constraints must refer to some defined parameter and value, and as we have seen in all the previous formats, auxiliary variables cannot be represented using ACTS.

The ACTS format also has additional features, such as support for mixed interaction strength (where the user can specify different strengths for different subsets of parameters).

In Figure~\ref{fig:acts-example} we show Example~\ref{ex:sut} represented using the ACTS format.

\begin{figure}
\centering
\begin{verbatim}
[System]
Name: MySUT
[Parameter]
OS (enum) : L,W,M,i,A
Pl (enum) : F,S,C,A
Re (enum) : K,F,H,W
Or (enum) : P,L
[Constraint]
C1: (OS = "L" || OS = "W" || OS = "M") =>
    (Or = "L" && Pl != "A")
C2: Pl = "S" => (OS = "M" || OS = "i")
C3: (OS = "i" || OS = "A") => Re != "K"
\end{verbatim}
\caption{Representation of SUT in Example~\ref{ex:sut} in ACTS format}
\label{fig:acts-example}
\end{figure}

\subsubsection{The new \emph{Extended} ACTS format}\label{sec:new-sut-format}

Here, we propose an extension of the ACTS format (see Section~\ref{sec:acts-format}) to describe SUTs with constraints.

As we have seen in Section~\ref{sec:avail-formats-suts}, none of the reviewed formats supports the addition of \emph{auxiliary variables} to the SUT constraints.
In other words, all the symbols that appear in the SUT constraints must refer to some parameter's value in the SUT.
This can be an important limitation of the current formats, as it will be more convenient to represent certain constraints using auxiliary variables (see Section~\ref{sec:sut-sat-gen}).

Essentially, we propose a simple extension to the ACTS format described in Section~\ref{sec:acts-format}.
In particular, we define a new section on the format under the tag \verb@[Auxiliar]@ where the user can define the list of auxiliary variables that will appear in the set of constraints.
Auxiliary variables are defined using the same syntax as parameters and can be of any of their types (see Section~\ref{sec:acts-format}).
Then, both the parameters and the auxiliary variables can be freely used in the SUT constraints.

In the next section, we present the \emph{SUT-G} generator, which uses the \emph{Extended ACTS} format to represent the generated SUT constraints.


\subsection{The \emph{SUT-G} Generator}\label{sec:sut-sat-gen}

The \emph{SUT-G} generator that we present in this section generates a SUT instance taking as input a SAT instance. Therefore, we first present several definitions regarding the SATisfiability (SAT) problem \cite{DBLP:series/faia/336}.

\begin{definition} A literal is a propositional variable~$x$ or a negated propositional variable~$\neg x$. A clause is a disjunction of literals. A Conjunctive Normal Form (CNF) formula, also called SAT instance, is a conjunction of clauses.  
\end{definition}

\begin{definition} A truth assignment for an instance $\varphi$ is a mapping that assigns to each propositional variable in $\varphi$ either 0 (False) or 1 (True). A truth assignment is \emph{partial} if the mapping is not defined for all the propositional variables in $\varphi$.
\end{definition}

\begin{definition}
A truth assignment $I$ satisfies a literal $x$ $(\neg x)$ if $I$ maps $x$ to 1 (0); otherwise, it is falsified. A truth assignment $I$ satisfies a clause if $I$ satisfies at least one of its literals; otherwise, it is violated or falsified.
Given a partial truth assignment $I$, a literal or a clause is undefined if it is neither satisfied nor falsified.  A clause $c$ is a unit clause under $I$ if $c$ is not satisfied by $I$ and contains exactly one undefined literal.  
\end{definition}

\begin{definition}
A SAT instance $\varphi$ is satisfiable if there is a truth assignment $I$, called model, such that $I$ satisfies all the clauses in $\varphi$. Otherwise, $\varphi$ is unsatisfiable. The Satisfiability (SAT) problem SAT problem is to determine whether a SAT instance is satisfiable.
\end{definition}

\begin{definition}
We refer as SAT solver to the implementation of an algorithm that takes as input a SAT instance and decides the SAT problem. The solver is said to be incremental if the input SAT instance can be modified and solved again while reusing some information from previous steps.
\end{definition}

\begin{definition}
Given a SAT instance $\varphi$ and a partial truth assignment $I$, we refer as Unit Propagation, denoted by $UP(I,\varphi)$, to the Boolean inference mechanism (propagator) defined as follows:  Find a unit clause in $\varphi$ under $I$, where $l$ is the undefined literal. Then, propagate the unit clause, i.e. extend $I$ with $x=1$ ($x=0$) if $l \equiv x$ ($l \equiv \neg x$) and repeat the process until a fixpoint is reached or a conflict is derived (i.e. a clause in $\varphi$ is falsified by $I$).\\[2mm]
We refer to $UP(I,\varphi)$ simply as $UP(\varphi)$ when $I$ is empty.
\end{definition}

Figure~\ref{fig:sat-interface} shows the interface of a modern incremental SAT solver.

\begin{figure}[ht]
    \centering
\begin{algorithmic}[1]

\STATE \COMMENT{\textbf{Attributes}}
\STATE $n\_vars$ \COMMENT{number of variables of the formula loaded}
\STATE $n\_conflicts$ \COMMENT{num. conflicts of the last call to $solve$}
\STATE $core$ \COMMENT{last core found}
\STATE $model$ \COMMENT{last model found}
\STATE
\STATE \COMMENT{\textbf{Methods}}
\STATE \textbf{function} $add\_clause(c: clause)$
\STATE \textbf{function} $solve(assumps: literals)$
\STATE \textbf{function} $propagate(assumps: literals)$
\STATE \textbf{function} $set\_seed(seed: int)$

\end{algorithmic}
    \caption{Interface of a modern incremental SAT solver}
    \label{fig:sat-interface}
\end{figure}

The input instance can be added to the SAT solver using $add\_clause$, which adds the specified clause to the solver.

The $solve$ method is one of the main methods of the SAT solver.
It returns SAT (UNSAT) if the input formula is satisfiable (unsatisfiable), and sets the variable $model$ ($core$) to the corresponding model (unsatisfiable core), as well as the variable $n\_conflicts$ to the number of conflicts produced in this call.
Additionally, the $solve$ method can receive a list of assumptions $assumps$ as a parameter, which will place an \emph{assumption} on the truth value of each of its literals before starting the solving process.

Regarding the $propagate$ method, it will apply Unit Propagation over the input instance, and can also receive a list of assumptions $assumps$ that it will \emph{assume} before.
This method returns the list of propagated literals.

Finally, we use the $set\_seed$ method to set the Random Number Generator seed of the SAT solver to the given $seed$.

All these methods will be used in the description of the \emph{SUT-G} generator.

As we mentioned in Section~\ref{sec:new-sut-format}, SUT-G will generate SUT models with constraints with auxiliary variables. Typically auxiliary variables are used to avoid the combinatorial explosion when translating constraints into CNF. For example, when we convert a Non-CNF circuit into CNF using the Tseitin transformation~\cite{Tseitin1983} that introduces fresh auxiliary variables. Also, auxiliary variables are used to translate into CNF well-known constraints, such as cardinality constraints or Pseudo-Boolean constraints~\cite{een_translating_2006}. Thanks to these variables, the translation is arc-consistent yet tractable in terms of size. Additionally, the introduction of auxiliary variables has been shown to be beneficial for SAT solvers in terms of boosting their performance~\cite{AnsoteguiM04}.

\begin{figure}[ht]
    \centering
\begin{algorithmic}[1]
    \REQUIRE{SAT formula $\varphi$, Number of SUT input parameters $n$, Max conflicts $c_{max}$, Min~conflicts~$c_{min}$, Assumptions increase $\Delta_\mathcal{A}$, Assumptions decrease $\nabla_\mathcal{A}$, Seeds $\mathcal{S}$, Maximum number of tries $MAX\_TRIES$}
    
    \ENSURE{ SUT model $S$}
    
    \STATE $\langle$status, $\mathcal{A} \rangle \leftarrow find\_satisfiable\_subproblem$( $\varphi$, $n$, $c_{max}$, $c_{min}$, $\Delta_\mathcal{A}$, $\nabla_\mathcal{A}$, $\mathcal{S}$, $MAX\_TRIES$ )\label{sutg:find-sat-subproblem}
    
    \IF{status $=$ FAIL}
        \RETURN{None}
    \ENDIF
    
    \STATE
    
    \STATE \COMMENT{Generate the constraints of the SUT}
    
    \FOR{$lit \in \mathcal{A}$}\label{sutg:iter-assumps}
        \STATE \COMMENT{Add each literal in $\mathcal{A}$ to $\varphi$ as unit clause}
        \STATE $\varphi \leftarrow \varphi \cup \{\{lit\}\}$\label{sutg:fix-assumps}
    \ENDFOR
    
    \STATE $\varphi' \leftarrow simplify(\varphi)$\label{sutg:simplify}
    
    \STATE \COMMENT{Select the input parameters of the SUT}
    
    \STATE $P \leftarrow sample(\varphi'.vars, n)$\label{sutg:select-params}
    
    \STATE $SUT \leftarrow \langle P,\varphi' \rangle$
    
    \RETURN{$SUT$}\label{gen:return}
    
\end{algorithmic}
    \caption{SUT Generation algorithm from SAT instances}
    \label{alg:sut-gen}
\end{figure}

Algorithm~\ref{alg:sut-gen} shows the pseudocode of the proposed \emph{SUT-G} generator. Essentially, this generator receives a SAT formula $\varphi$ and returns a SUT model with constraints, where this input formula $\varphi$ has been adapted to the desired difficulty.
We need a \emph{hardness measure} to adapt $\varphi$, and in our case, we decided to use the number of conflicts  the SAT solver needs on average to solve the instance.

Aside from $\varphi$, \emph{SUT-G} receives as input parameters the number of parameters $n$ of the output SUT, the maximum and minimum number of conflicts of the SUT constraints ($c_{max}$ and $c_{min}$), and other parameters that control several aspects of SUT generation that will be explained later ($\Delta_\mathcal{A}$, $\nabla_\mathcal{A}$, $\mathcal{S}$ and $MAX\_TRIES$).

On a high level, the \emph{SUT-G} generator will try to find a satisfiable subproblem over the input formula $\varphi$ (line~\ref{sutg:find-sat-subproblem}).
This subproblem will match all the requirements specified in the input parameters of the algorithm.
In case the input formula is unsatisfiable or if the requirements cannot be fulfilled, the \verb@find_satisfiable_subproblem@ method will return a \emph{fail} status and the generator will exit.
Otherwise, \verb@find_satisfiable_subproblem@ will return a \emph{success} status and a list of assumptions that will adapt the input formula to the hardness requirements when applied to the original formula $\varphi$.

To apply these modifications to the original formula, we just add as unit clauses each literal in $\mathcal{A}$ (lines~\ref{sutg:iter-assumps} and~\ref{sutg:fix-assumps}).
Then, we \emph{simplify} this formula to obtain $\varphi'$ (line~\ref{sutg:simplify}), which will become the SUT constraints.

This simplification process propagates all the unit clauses in the formula and obtains a list of literals $lits$ with a fixed value.
Then, it iterates all the clauses in $\varphi$ and eliminates all the ones that have a literal in $lits$ (i.e. this clause is satisfied and can be ignored in the resulting formula).
On the other hand, if a literal in a clause appears with opposite polarity in $lits$ we can remove this literal from the clause.
Finally, we rename the variables in the new formula to ensure that all the variables from $1$ to $\varphi'.n\_vars$ appear in the constraints.

To finish the generation of the SUT, we just have to select its parameters.
In this case, they are randomly selected from $1$ to $\varphi'.n\_vars$ taking into account the number of parameters $n$ specified by the user (line~\ref{sutg:select-params}).
The rest of the variables in $\varphi'$ will be auxiliary variables.

Figure~\ref{alg:sut-gen-aux} shows the auxiliary functions used by the \emph{SUT-G} generator.

\begin{figure}[ht]
    \centering
\begin{algorithmic}[1]
    \STATE \textbf{function} $find\_satisfiable\_subproblem$($\varphi$, $n$, $c_{max}$, $c_{min}$, $\Delta_\mathcal{A}$, $\nabla_\mathcal{A}$, $\mathcal{S}$, $MAX\_TRIES$)
    
        \STATE $sat \leftarrow$ incremental SAT solver initialized with $\varphi$ \label{find:init-sat}
    
        \STATE $\mathcal{A} \leftarrow \emptyset$\label{find:init-assumps}
        
        \STATE $\langle model, c\rangle \leftarrow solve\_subproblem(sat, \mathcal{A}, \mathcal{S})$\label{find:solve-1}
            
        \IF{$model = \emptyset$ \textbf{or} $sat.n\_vars - |sat.propagate(\mathcal{A})| < n$ \textbf{or} $c < c_{min}$}\label{find:check-varphi}
            \RETURN $\langle$ FAIL, \_ $\rangle$\label{find:fail-1}
        \ENDIF
        
        \STATE $tries \leftarrow 1$\label{find:loop-start}
        
        \WHILE{$tries < MAX\_TRIES$}
        
            \IF{$sat.n\_vars - |sat.propagate(\mathcal{A})|$ $< n$ \textbf{or} $c < c_{min}$}\label{find:check-easy}
                \IF{$|\mathcal{A}| = 0$}\label{find:empty-assumps}
                    \RETURN $\langle$ FAIL, \_ $\rangle$\label{find:fail-assumps}
                \ENDIF
                \STATE $\mathcal{A} \leftarrow \mathcal{A} \setminus sample$($\mathcal{A}$, $\nabla_\mathcal{A}$)\label{find:dec-assumps}
                
            \ELSIF{$c > c_{max}$}\label{find:check-hard}
                \STATE $\mathcal{A} \leftarrow \mathcal{A} \cup sample(model \setminus \mathcal{A}, \Delta_\mathcal{A})$\label{gen:sample-cmax}
            
            \ELSE
                \RETURN $\langle$ SUCCESS, $sat.propagate(\mathcal{A})$ $\rangle$\label{find:success}
            \ENDIF
            
            \IF{$sat.n\_vars - |sat.propagate(\mathcal{A})| \geq n$}\label{find:check-vars}
            
                \STATE $\langle model, c\rangle \leftarrow solve\_subproblem(sat, \mathcal{A}, \mathcal{S})$\label{find:solve-2}
                
                \STATE $tries \leftarrow tries + 1 $\label{gen:adapt-end}\label{find:loop-end}
            \ENDIF
        \ENDWHILE
        
        \RETURN $\langle$ FAIL, \_ $\rangle$\label{find:fail-2}
    \STATE \textbf{end function}
    
    \STATE

    \STATE \textbf{function} $solve\_subproblem(sat, \mathcal{A}, \mathcal{S})$\label{solve:def}
        \STATE $c \leftarrow 0$\label{solve:init-c}
        
        \STATE $models \leftarrow \emptyset$\label{solve:init-models}
        
        \FOR{$seed \in \mathcal{S}$}\label{solve:iter-seeds}
            \STATE $sat.set\_seed(seed)$\label{solve:set-seed}
            
            \IF{$sat.solve(\mathcal{A}) = UNSAT$}\label{solve:solve}
                \RETURN $\langle$ \_, $\emptyset$, \_ $\rangle$\label{solve:fail}
            \ENDIF
            
            \STATE $models \leftarrow models \cup \{sat.model\}$\label{solve:add-model}
            
            \STATE $c \leftarrow c + sat.n\_conflicts$\label{solve:sum-confl}
        \ENDFOR
        
        \RETURN{$\langle sample(models, 1), c / |\mathcal{S}| \rangle$}\label{solve:return}
    \STATE \textbf{end function}

\end{algorithmic}
    \caption{Auxiliary functions for SUT-Gen algorithm}
    \label{alg:sut-gen-aux}
\end{figure}

Function \verb@find_satisfiable_subproblem@ (used in line~\ref{sutg:find-sat-subproblem} of Figure~\ref{alg:sut-gen}) starts initializing an incremental SAT solver with the input formula $\varphi$, as well as the set of assumptions $\mathcal{A}$ to the empty set (lines~\ref{find:init-sat} and~\ref{find:init-assumps}).
Then, it uses the function \verb@solve_subproblem@ to find the \emph{hardness} of the subproblem in terms of conflicts $c$, which is defined by the original formula $\varphi$ and the set of assumptions $\mathcal{A}$\footnote{Initially $\mathcal{A}$ is empty, this is equivalent to solving the original formula $\varphi$} (line~\ref{find:solve-1}).
Additionally, \verb@solve_subproblem@ will return a model that will be used later.

In case the original formula is UNSAT (which we represent when the $model$ is empty), or there are not enough \emph{unfixed} variables that could become parameters of the SUT\footnote{We are also considering as \emph{fixed} variables that cannot become SUT parameters the Unit Propagation of $\mathcal{A}$. If some of these variables is selected as a parameter of the SUT it will have just one possible value.} ($sat.n\_vars - |sat.propagate(\mathcal{A})| < n$), or the formula is too easy ($c < c_{min}$), the function will return a \emph{fail} status (lines~\ref{find:check-varphi} and~\ref{find:fail-1}).

Lines~\ref{find:loop-start}~-~\ref{find:loop-end} show the main loop of the function.
We consider a maximum number of tries $MAX\_TRIES$ to obtain the desired subproblem (which is a parameter of \emph{SUT-G}, see Figure~\ref{alg:sut-gen}).
In case these $MAX\_TRIES$ are reached and the desired subproblem could not be obtained, we also return a \emph{fail} status, as shown in line~\ref{find:fail-2}).

At each iteration of the main loop, the algorithm tries to find a subproblem by increasing or decreasing the assumptions $\mathcal{A}$ that will contain enough variables (to become the parameters in the SUT) and with the desired hardness in terms of the number of conflicts.

The idea is that by increasing the assumptions we are making the problem \emph{easier} and decreasing them \emph{harder}.
This is however a greedy approach, and there might be the case where the problem becomes \emph{harder} when increasing the constraints, and vice-versa.
Nonetheless, we found this greedy process quite successful when generating interesting SUTs, as we show in our experimental evaluation (see Section~\ref{sec:exps-ipog-bot}).

In line~\ref{find:check-easy} we check if there are not enough \emph{unfixed} variables that could become parameters of the SUT ($sat.n\_vars - |sat.propagate(\mathcal{A})| < n$), or the current subproblem is too easy ($c < c_{min}$).
If that is the case, we try to decrease the assumptions $\mathcal{A}$.
In case we have some literals in the assumptions, we remove $\nabla_\mathcal{A}$ random literals in $\mathcal{A}$ in line~\ref{find:dec-assumps}, which is one of the input parameters of \emph{SUT-G} (see Figure~\ref{alg:sut-gen}).

However, it might be the case where the assumptions that we have at this point are empty (line~\ref{find:empty-assumps}).
This means that we cannot simplify the subproblem anymore, and therefore we return a \emph{fail} (line~\ref{find:fail-assumps}).
Notice that this case is possible due to the incremental nature of the SAT solver.
At first, we might find that the original formula is too hard in line~\ref{find:solve-1}, so in the main loop, we will increase the assumptions (as it will be explained next).
Then, the SAT solver might \emph{learn} some clauses that make easier the original problem\footnote{These clauses are kept between calls to $solve$}.
At this point, decreasing assumptions will not help to increase the hardness of the subproblem, so when we reach $|\mathcal{A}| = 0$ we can stop the loop.

On the other hand, if the current subproblem is too hard ($c > c_{max}$, line~\ref{find:check-hard}), we increase the assumptions $\mathcal{A}$ by adding to them $\Delta_\mathcal{A}$ random literals of the $model$ returned by \verb@solve_subproblem@. ($\Delta_\mathcal{A}$ is another input parameter of \emph{SUT-G}, see Figure~\ref{alg:sut-gen}).

If none of these cases is found, we are within the conflicts range and there are enough \emph{unfixed} variables that can become parameters of the output SUT.
Therefore, we just return a \emph{success} state and the list of \emph{propagated} assumptions (line~\ref{find:success}).

Finally, in lines~\ref{find:solve-2} and~\ref{find:loop-end} we call \verb@solve_subproblem@ again to update the number of conflicts $c$ and the model according to the changes in $\mathcal{A}$ that we performed, and we add one try to $tries$.
Notice however that these two lines are only executed in case there are enough \emph{unfixed} variables in the subproblem (line~\ref{find:check-vars}).
If this is not the case, we will keep decreasing $\mathcal{A}$ without solving the subproblem nor increasing the number of tries.

In line~\ref{solve:def} we describe function \verb@solve_subproblem@.
This function will solve each subproblem and estimate its hardness in terms of conflicts.

First, it starts initializing the number of conflicts $c$ and a set of models $models$ to 0 and $\emptyset$ respectively (lines~\ref{solve:init-c} and~\ref{solve:init-models}).

Then, in line~\ref{solve:iter-seeds} it iterates $\mathcal{S}$, which contains a list of seeds provided by the user as input in \emph{SUT-G}.
We set the Random Number Generator of the SAT solver to the current $seed$ in line~\ref{solve:set-seed}, and proceed to solve the current subproblem in line~\ref{solve:solve} (which is the formula in the SAT solver plus $\mathcal{A}$).
In this same line, we check if the subproblem is UNSAT, and return an empty model in line~\ref{solve:fail} if this is the case.
Otherwise, when the subproblem is SAT, we retrieve the model and add it to $models$ in line~\ref{solve:add-model}.
Additionally, we compute the number of conflicts and sum them to $c$ in line~\ref{solve:sum-confl}.
At the end in line~\ref{solve:return} we return one random model from the list of models, as well as the average on the number of conflicts~\footnote{We return the average on the number of conflicts over all the $\mathcal{S}$ to mitigate lucky and unlucky runs of the SAT solver}.

For this version of the algorithm, we are considering that all the parameters in the SUT have domain 2. In next versions, we will consider the generation of SUT models with mixed domains.

Figure~\ref{fig:sut-g-diagram} shows a summary diagram of the SUT-G generator approach.

\begin{figure}[ht]
\centering
\begin{tikzpicture}[node distance=1.5cm]
    \node (n1) [rect] {SAT instance $\varphi$};
    \node (n2) [rect, text width=5cm, below of = n1] {Find a satisfiable subproblem $\varphi'$ from $\varphi$ with at least $k$ variables and of at least some \emph{hardness}};
    \node (n3) [rect, text width=5cm, below of = n2] {Let the SUT constraints be the clauses in $\varphi'$};
    \node (n4) [rect, text width=5cm, below of = n3] {Let the SUT parameters $P$ be a sample of $k$ variables from $\varphi'$};
    \node (n5) [rect, below of = n4] {Build SUT $\langle P, \varphi' \rangle$};
    
    \draw [->, thick] (n1) -- (n2);
    \draw [->, thick] (n2) -- (n3);
    \draw [->, thick] (n3) -- (n4);
    \draw [->, thick] (n4) -- (n5);
\end{tikzpicture}
\caption{Summary diagram of the SUT-G generator}
\label{fig:sut-g-diagram}
\end{figure}

First, we receive a SAT instance $\varphi$ as input.
Then, we find a satisfiable subproblem $\varphi'$ over this SAT instance $\varphi$ which must have at least \emph{k} variables (which will became the SUT parameters) and that fulfils some \emph{hardness} requirements (in terms of minimum and maximum number of conflicts).
In case this subproblem cannot be found, we would return a failure for the input SAT instance.
Otherwise, we take all the clauses of the subproblem as the SUT constraints and we sample \emph{k} variables from $\varphi'$ as the SUT parameters (the remaining variables are considered auxiliary variables).
Finally, we build the output SUT with these two parts.


\section{Algorithms for Building MCACs}
\label{sec:algs}

In Section~\ref{sec:gen-bench-sat} we described the \emph{SUT-Gen} Algorithm, which can generate crafted SUTs with constraints for CT. Notice that these SUTs can be potentially used to evaluate any CT algorithm. Nonetheless, in the remaining of this work, we will focus on the generation of Mixed Covering Arrays With Constraints (MCACs), one of the most extended CT methods.

In this section, we will review two of the main state-of-the-art algorithms for building MCACs greedily, which are the In-Parameter-Order-General (IPOG) algorithm and the Build-One-Test Iterative-Test-Suite (BOT-its) algorithm and its variations. Then, in Section~\ref{sec:sota-mcac-exps}, we will evaluate its performance on a set of benchmarks generated with the \emph{SUT-Gen} algorithm.

First of all, we will start by introducing some definitions regarding MCACs.

\begin{definition} An assignment is a set of pairs $(p,v)$ where $p$ is a variable and $v$ is a value of the domain of $p$. A test case for $S$ is a full assignment $A$ to the variables in $S_P$ such that $A$ entails $S_\varphi$  (i.e. $A \models S_\varphi$). A parameter tuple of $S$ is a subset $\pi \subseteq S_P$. A value tuple of $S$ is a partial assignment to $S_P$; in particular, we refer to a value tuple of length $t$ as a $t$-tuple.
\end{definition}

\begin{example}
A test case for the SUT model in Example \ref{ex:sut} is $\{(OS, W), (Pl, F), (Re, K), (Or, L)\}$.

\noindent $\{Os, Pl\}$ is a parameter tuple and $\{(OS, W), (Pl, F)\}$ is a $2$-tuple (i.e. a value tuple for strength $t=2$).
\end{example}

\begin{definition}
A $t$-tuple $\tau$ is forbidden if $\tau$ does not entail $S_\varphi$ (i.e. $\tau \models \neg S_\varphi$). Otherwise, it is allowed. We refer to the set of allowed $t$-tuples as $\Tau_a = \{\tau \mid \tau \not\models \neg S_\varphi\}$.
\end{definition}

\begin{example}
The $2$-tuple $\{(OS, i), (Re, K)\}$ is a forbidden tuple. There are 69 allowed $2$-tuples $\lvert\Tau_a\rvert$ in the SUT model of Example \ref{ex:sut}.
\end{example}

\begin{definition} A test case $\upsilon$ \emph{covers} a value tuple $\tau$ if both assign the same domain value to the variables in the value tuple, i.e., $\upsilon \models \tau$. A test suite $\Upsilon$ \emph{covers} a value tuple $\tau$ (i.e., $\tau \subseteq \Upsilon$) if there exist a test case $\upsilon \in \Upsilon$ s.t. $\upsilon \models \tau$. We refer to $\upsilon \not\models \tau$ ($\tau \not\subseteq \Upsilon$) when a test case (test suite) \emph{does not cover} $\tau$.
\end{definition}

\begin{example}
The test case $\upsilon = \{(OS, W), (Pl, F), (Re, K), (Or, L)\}$ covers the following $t$-tuples for $t=2$:

\noindent$\{(OS, W), (Pl, F)\},\{(OS, W), (Re, K)\},\{(OS, W), (Or, L)\},$ \\
\noindent$\{(Pl, F), (Re, K)\}, \{(Pl, F), (Or, L)\}, \{(Re, K), (Or, L)\}$
\end{example}

\begin{definition}
A Mixed Covering Array with Constraints (MCAC), denoted by $CA(N;t,S)$, is a set of $N$ test cases for a SUT model $S$ such that all $t$-tuples are at least covered by one test case. The term \emph{Mixed} reflects that the domains of the parameters in $S_P$ are allowed to have different cardinalities. The term \emph{Constraints} reflects that $S_\varphi$ is not empty.
\end{definition}

\begin{example}
Table \ref{tab:ca} shows an MCAC for the SUT $S$ in Example \ref{ex:sut}
\end{example}

\begin{definition}
The MCAC problem is to find an MCAC of size $N$.
\end{definition}

\begin{definition}
The Covering Array Number, $CAN(t,S)$, is the minimum $N$ for which there exists an MCAC $CA(N;t,S)$. 
The Covering Array Number problem is to find an MCAC of size $CAN(t,S)$.
\end{definition}

\begin{example}
Table \ref{tab:can} shows an optimal $CA(N;t,S)$ for the SUT model in Example \ref{ex:sut}. In this case $CAN(2,S)=21$.

\begin{table}
\parbox{.45\linewidth}{
\centering
\scalebox{0.7}{
\begin{tabular}{c|c|c|c|c}
    test & OS & Pl & Re & Or \\
    \hline
    $\upsilon_{1}$ & L & F & F & L \\
    $\upsilon_{2}$ & L & C & H & L \\
    $\upsilon_{3}$ & W & F & W & L \\
    $\upsilon_{4}$ & W & C & K & L \\
    $\upsilon_{5}$ & M & F & H & L \\
    $\upsilon_{6}$ & M & S & W & L \\
    $\upsilon_{7}$ & M & C & F & L \\
    $\upsilon_{8}$ & i & F & H & P \\
    $\upsilon_{9}$ & i & S & F & P \\
    $\upsilon_{10}$ & i & C & W & P \\
    $\upsilon_{11}$ & i & A & F & L \\
    $\upsilon_{12}$ & A & F & H & P \\
    $\upsilon_{13}$ & A & C & W & L \\
    $\upsilon_{14}$ & A & A & F & P \\
    $\upsilon_{15}$ & L & F & K & L \\
    $\upsilon_{16}$ & M & S & K & L \\
    $\upsilon_{17}$ & i & S & H & L \\
    $\upsilon_{18}$ & A & A & H & L \\
    $\upsilon_{19}$ & i & A & W & L \\
    $\upsilon_{20}$ & W & F & F & L \\
    $\upsilon_{21}$ & W & F & H & L \\
    $\upsilon_{22}$ & L & F & W & L
\end{tabular}
}
\caption{$CA(22;2,S)$ for the SUT in example \ref{ex:sut}}
\label{tab:ca}
}
\hfill
\parbox{.45\linewidth}{
\centering
\scalebox{0.7}{
\begin{tabular}{c|c|c|c|c}
    test & OS & Pl & Re & Or \\
    \hline
    $\upsilon_{1}$ & L & C & K & L \\
    $\upsilon_{2}$ & L & F & F & L \\
    $\upsilon_{3}$ & L & C & H & L \\
    $\upsilon_{4}$ & L & C & W & L \\
    $\upsilon_{5}$ & W & F & K & L \\
    $\upsilon_{6}$ & W & C & F & L \\
    $\upsilon_{7}$ & W & F & H & L \\
    $\upsilon_{8}$ & W & F & W & L \\
    $\upsilon_{9}$ & M & S & K & L \\
    $\upsilon_{10}$ & M & S & F & L \\
    $\upsilon_{11}$ & M & C & H & L \\
    $\upsilon_{12}$ & M & F & W & L \\
    $\upsilon_{13}$ & i & C & F & P \\
    $\upsilon_{14}$ & i & S & H & P \\
    $\upsilon_{15}$ & i & S & W & P \\
    $\upsilon_{16}$ & A & A & F & L \\
    $\upsilon_{17}$ & A & A & H & P \\
    $\upsilon_{18}$ & A & F & W & P \\
    $\upsilon_{19}$ & i & A & W & L \\
    $\upsilon_{20}$ & A & C & H & L \\
    $\upsilon_{21}$ & i & F & H & L
\end{tabular}
}
\caption{$CA(21;2,S)$ for the SUT in example \ref{ex:sut}. This corresponds to the $CAN(2,S)$.}
\label{tab:can}
}
\end{table}
\end{example}

\subsection{The In-Parameter-Order-General (IPOG) Algorithm}\label{sec:ipog}

The In-Parameter-Order-General (IPOG) algorithm \cite{https://doi.org/10.1002/stvr.381} provides an incomplete approach for solving the $CAN(t,S)$ problem. Figure \ref{fig:ipog} shows its pseudocode.

\begin{figure}[ht]
\centering
\begin{algorithmic}[1]
    \REQUIRE{SUT model $S$, strength $t$}
    
    \ENSURE{Test suite $\Upsilon$}
    
    \STATE $P \leftarrow argsort_{p \in S_p} d(p)$\label{ipog:sort-params}
    
    \STATE $\Upsilon \leftarrow CAN(t, \langle \{P_1, ..., P_t\}, S_\varphi \rangle)$\label{ipog:first-can}
    
    \FOR{$p$ \textbf{in} $P_{t+1}\ ...\ P_{\lvert P \rvert}$}\label{ipog:main-loop}
        \STATE \COMMENT{Initialize tuples pool}
        
        \STATE $\rho \leftarrow \{\tau\ \lvert\ \tau \in d(p)$ combined with all the tuples of size $t - 1$ in $\Upsilon\}$ \label{ipog:init-pool}
        
        \STATE \COMMENT{Horizontal extension}
        \FOR{$\upsilon$ \textbf{in} $\Upsilon$}\label{ipog:hori-start}
            \STATE Choose best v $\in d(p)$ s.t. $\upsilon \cup \{(p, $v$)\}$ covers more tuples in $\rho$ and is consistent with $S_\varphi$\label{ipog:best-v}
            
            \STATE $\upsilon \leftarrow \upsilon \cup \{(p,$v$)\}$
            
            \STATE $\rho_v \leftarrow \{\tau\ \lvert\ (\tau \in \rho) \land (\upsilon \models \tau)\}$
            
            \STATE $\rho \leftarrow \rho \setminus \rho_v$
            
            \IF{$\lvert \rho \rvert = \emptyset$}
                \STATE \textbf{break}\label{ipog:hori-end}
            \ENDIF
        \ENDFOR
        
        \IF{$\lvert \rho \rvert \neq \emptyset$}
            \FOR{$\tau\ \lvert\ (\tau \in \rho)\ \land$ $(\tau$ consistent with $S_\varphi)$}\label{ipog:loop-tuples}
                \STATE \COMMENT{Fill the empties}
                
                \STATE covered $\leftarrow$ false \label{ipog:fill-start}
                
                \FOR{$\upsilon$ \textbf{in} $\Upsilon$}
                    \IF{$\tau \cup \upsilon$ is possible and consistent with $S_\varphi$}\label{ipog:consistent-tau-upsilon}
                        \STATE $\upsilon \leftarrow \upsilon \cup \tau$
                        
                        \STATE covered $\leftarrow$ true
                        
                        \STATE \textbf{break}\label{ipog:fill-end}
                    \ENDIF
                \ENDFOR
                
                \STATE \COMMENT{Vertical growth}
                \IF{covered = false}
                    \STATE $\Upsilon \leftarrow \Upsilon \cup \{\upsilon\ s.t.\ \upsilon \models \tau\}$\label{ipog:vert}
                \ENDIF
            \ENDFOR
        \ENDIF
    \ENDFOR
    
    \RETURN{$\Upsilon$}
    
\end{algorithmic}
\caption{IPOG algorithm}
\label{fig:ipog}
\end{figure}

Essentially, the IPOG algorithm builds an MCAC of strength $t$ by iteratively adding one parameter to the currently built MCAC.
It starts sorting the parameters $S_p$ of the input SUT according to their domain in descending order (line~\ref{ipog:sort-params}).
The first test suite $\Upsilon$ that it builds is the one that contains a test case with just each valid $t$-tuple for the first $t$ parameters (line~\ref{ipog:first-can}).
The rest of the parameters will have an \emph{empty} value.
Then, it iteratively adds one parameter to the current test suite (line~\ref{ipog:main-loop}).
The main idea is that, for each new parameter $p$, we only have to consider the $t$-tuples of this parameter with the previous ones.
This way we are reducing the number of $t$-tuples that we have to keep track of.

First of all, the IPOG algorithm will try to cover as many tuples as possible for each of the existing test cases in $\Upsilon$ by fixing a value v for the parameter $p$ (lines~\ref{ipog:hori-start}-~\ref{ipog:hori-end}, what we call \emph{horizontal extension}).
If no tuples can be covered for a given test case $\upsilon$, we will keep the value for parameter $p$ in $\upsilon$ as \emph{empty}.

In case there are tuples that were not covered in the \emph{horizontal extension} procedure, the IPOG algorithm will try to fit each of the valid tuples in some of the current test cases in $\Upsilon$ taking advantage of the \emph{empty} values of each test (lines~\ref{ipog:fill-start}-~\ref{ipog:fill-end}, what we call \emph{fill the empties}).
Otherwise, a new test $\upsilon$ with just the current tuple is created and added to the current test suite $\Upsilon$ (line~\ref{ipog:vert}, what we call \emph{vertical growth}).
Notice also that the rest of the parameters will have an \emph{empty} value.

The algorithm proceeds until all the parameters in $S_p$ have been processed.

This version of the IPOG algorithm uses a SAT solver to handle the SUT constraints $S_\varphi$.
Although there are other alternative approaches to handle SUT constraints in the IPOG algorithm~\cite{7107441}, these are only suitable for simpler constraints where auxiliary variables are not present and therefore not considered for this work.
Notice that every time that the IPOG algorithm fixes a value or a tuple in a test case, it must ensure that it is consistent with the SUT constraints $S_\varphi$, as these values will be used for the future parameters (see lines~\ref{ipog:best-v}, \ref{ipog:loop-tuples} and \ref{ipog:consistent-tau-upsilon}).

In the next section, we review the BOT-its algorithm and its variants~\cite{DBLP:conf/cp/AnsoteguiOT21}, which is based in~\cite{Yamada16}.


\subsection{The Build-One-Test (BOT-its)/\emph{imply} Algorithm}\label{sec:bot}

Another state-of-the-art approach for building MCACs is the OTAT framework~\cite{BryceCC05}.
The algorithm starts with an empty test suite $\Upsilon$ and iteratively adds one test case $\upsilon$ at a time such that $\upsilon$ covers more tuples.
This process is repeated until all the valid $t$-tuples for the input SUT and strength $t$ are covered.

There are several implementations for the OTAT framework, being PICT~\cite{Czerwonka06} one of them.
Another implementation is described in~\cite{Yamada16}, where the authors propose a variant of OTAT that can handle SUT constraints more efficiently by using SAT technology.
In~\cite{DBLP:conf/cp/AnsoteguiOT21}, the authors describe the BOT-its algorithm and its variants.
These algorithms are extensions of the work in~\cite{Yamada16} that allow i) the refinement of the working test suites to reduce their size using MaxSAT technology through the RBOT-its algorithm; ii) to scale the BOT-its algorithm to higher strengths by limiting the pool of $t$-tuples in the PBOT-its; and iii) the combination of the RBOT-its and PBOT-its algorithms into the PRBOT-its algorithm.

Figure~\ref{alg:botits} shows the pseudocode for the basic BOT-its algorithm. For a more detailed description of this algorithm, as well as its variants, see~\cite{DBLP:conf/sat/AnsoteguiOPPST21,Yamada16}.

\begin{figure}[ht]
    \centering
\begin{algorithmic}[1]
    \REQUIRE{SUT model $S$, strength $t$, consistency check conflict budget $cb$}
    
    \ENSURE{Test suite $\Upsilon$}
    
    \STATE$\Upsilon \leftarrow \emptyset$ \COMMENT{Working test suite}\label{botits:init-ts}
    
   \STATE $\rho \leftarrow$ pool with all $t$-tuples of $S$
    
    \STATE$sat \leftarrow$ incremental SAT solver initialized with $S_\varphi$\label{botits:init-sat}
    
    \WHILE{$\rho \neq \emptyset$}
        \STATE $\upsilon, \rho \leftarrow BOT(S, \rho, sat, cb)$
        
        \STATE $\Upsilon \leftarrow \Upsilon \cup \{\upsilon\}$
        
        \STATE $\rho_{\upsilon} \leftarrow \{\tau \mid \tau \in \rho \wedge \upsilon \models \tau\}$ \COMMENT{Tuples in $\rho$ covered by $\upsilon$}\label{botits:tupl-remove-start}
        
        \STATE$\rho \leftarrow \rho \setminus \rho_{\upsilon}$
        
    \ENDWHILE
    
    \RETURN{$\Upsilon$}
    
\end{algorithmic}
    \caption{Build One Test - Incremental Test Suite algorithm (BOT-its)} \label{alg:botits}
\end{figure}

This algorithm builds an MCAC by iteratively calling Algorithm BOT (Build One Test), which greedily builds a new test case. Initially, BOT-its initializes the working test suite $\Upsilon$ to empty, the pool of tuples $\rho$ that need to be covered to all the $t$-tuples in the input SUT $S$, and an incremental SAT solver initialized with $S_\varphi$ (lines~\ref{botits:init-ts}~-~\ref{botits:init-sat}). Then, it extends the working test suite $\Upsilon$ incrementally by adding a new test case $\upsilon$ computed by the BOT algorithm, and the $t$-tuples that are covered by $\upsilon$ are removed from $\rho$.
Finally, the algorithm returns the test suite when the pool of tuples $\rho$ becomes empty (at this point the test suite is indeed an MCAC).

The main difference between the BOT-its algorithm with other algorithms that also follow the OTAT framework is the management of the SUT constraints. In this case, it allows us to build a test case that is not consistent with the SUT constraints while building it, and it \emph{amends} it once the test case is completed. This way it can skip most of the queries to the SAT solver at the expense of slightly increasing the test suite size. Notice that, for example, the IPOG algorithm cannot apply this same procedure (see Section~\ref{sec:ipog}). This algorithm is based on Algorithm 5 in~\cite{Yamada16}.

Another difference between the OTAT-based algorithm and IPOG is that the first ones allow \emph{online} testing (i.e. applying each test to the SUT as soon as they are generated), while for IPOG it is required to wait until the algorithm finishes.
Although this feature can be useful for certain scenarios, it comes with the drawback of higher memory consumption (OTAT-based algorithms need to keep in memory all the $t$-tuples, while IPOG only the ones that involve the current parameter with the previous ones).
A way to mitigate this issue is presented in the PBOT-its algorithm~\cite{DBLP:conf/sat/AnsoteguiOPPST21}.


\section{Assessing MCAC Tools with \emph{SUT-Gen}}
\label{sec:sota-mcac-exps}

In this section, we assess the quality of the SUT constraints generated by SUT-G.
In our experimental evaluation, we will show (by empirical evaluation) that the SAT-based SUTs we generate are different from the random SUTs available in the CT competition (even using the same number of parameters) both in terms of runtime and size of the covering array.
This experimental evaluation has three main goals:

\begin{enumerate}
    \item Compare the SUT-G generator with other state-of-the-art random SUT generators.
    \item Analyze the impact of the generated SUT constraints with respect to the same unconstrained SUT.
    \item Evaluate and compare the differences in test suite size and runtime of the mentioned MCAC algorithms using the generated CT benchmarks, and compare their behaviour with the available CT benchmarks.
\end{enumerate}

Unfortunately, we did not find any state-of-the-art tool for generating MCACs that supported the \emph{Extended ACTS} format described in Section~\ref{sec:new-sut-format}.
Although there are some MCAC tools that could be adapted to support this format (one of them is the ACTS tool~\cite{BorazjanyYLKK12}, in particular, their IPOG implementation that uses a Constraint Programming (CP) solver to handle SUT constraints~\cite{6569736}), these modifications were out of the scope of this work.
Instead of that, we decided to adapt our own version of the \emph{IPOG} algorithm (named \emph{CTLog}) with full support for both the \emph{ACTS} and \emph{Extended ACTS} SUT formats.
To our best knowledge, the main difference between these two implementations is that in CTLog constraints are handled using the SAT solver Glucose 4.1~\cite{audemard2013improving}, whereas in ACTS they are handled using a CP solver.
Additionally, ACTS is implemented in the Java programming language, and our version is implemented in Python and Nim.
Also, there are some details on how ties are broken in the implementation of ACTS, which we have observed has a great impact. 

To ensure that our implementation is competitive with ACTS, we tested both algorithms using 58 state-of-the-art benchmarks~\cite{cohen2008constructing,segall11,Yu15,Yamada16} for strengths $t=\{2,3,4\}$, as both implementations support this input format.
Additionally, we included the results of the \emph{BOT-its} algorithm. We will use them as a baseline in future comparisons.
Table~\ref{tab:acts-vs-ctlog} shows these results. We executed the ACTS implementation one time and the algorithms in CTLog 5 times with different seeds to mitigate the stochastic behaviour of the algorithm\footnote{The ACTS tool does not expose the \emph{seed} parameter}.
We report the average test suite size and run time of the different algorithms

As we can observe, the \emph{CTLog} implementation of IPOG is able to solve 9 more instances than \emph{ACTS} for strengths $t=\{2,3,4\}$. In general, our IPOG implementation finds MCACs faster than \emph{ACTS}. Regarding MCAC sizes, both approaches obtain similar results. Therefore, it makes sense to use our own IPOG implementation as we have demonstrated that it is equivalent to or even better than the \emph{ACTS} implementation.

We selected the \emph{BOT-its} and \emph{PBOT-its}~\cite{DBLP:conf/cp/AnsoteguiOT21} algorithms to compare against IPOG (Sections~\ref{sec:ipog} and~\ref{sec:bot}). All our algorithms are implemented in Python except for the most critical parts, which have been implemented in Nim\footnote{The Nim programming language (\url{https://nim-lang.org/})}. We use the OptiLog~\cite{DBLP:conf/sat/AlosAST22} Python framework to efficiently use SAT solvers and encoders. In both cases, we use the Glucose 4.1 SAT solver~\cite{audemard2013improving}. These algorithms have been adapted to support the new SUT format discussed in Section~\ref{sec:new-sut-format}.

Table~\ref{tab:acts-vs-ctlog} shows that, in general, BOT-its obtains worse run times than the IPOG implementation in CTLog.
We will observe how this behaviour is modified when using the benchmarks generated with SUT-G.

Regarding the generated benchmarks, we decided to use industrial SAT instances for our study.
The main idea is that the generated SUT constraints will keep some of the industrial-like structure of the original SAT formula, and can be a good approximation of the constraints of other real-world industrial SUTs.

We selected the 165 instances of the SAT competition 2009~\cite{SAT-competition-2009} to try to find solvable satisfiable instances for modern SAT solvers more easily\footnote{Theoretically, SAT solvers that apply modern techniques should be able to solve more efficiently instances of previous SAT competitions.}. Notice that this selection is only for convenience, as \emph{SUT-G} is able to adapt the hardness of the input instance as explained in Section~\ref{sec:gen-bench-sat}.
From all the instances, we could only consider those that were satisfiable and complied with the conflicts threshold that we describe next.

We also used Glucose 4.1~\cite{audemard2013improving} as incremental SAT solver for Algorithm~\ref{alg:sut-gen}. We set the parameters \texttt{rnd-freq} and \texttt{rnd-pol-freq} of the solver to 0.5 to allow some variations on the results of the \verb@solve_subproblem@ function. 

\begin{table*}
\centering
\resizebox{.8\textwidth}{!}{
\begin{tabular}{l|rrrrrr|rrrrrr|rrrrrr}
\toprule
{} & \multicolumn{6}{c|}{$t=2$} & \multicolumn{6}{c|}{$t=3$} & \multicolumn{6}{c}{$t=4$} \\
\midrule
{} & \multicolumn{2}{c}{ACTS} & \multicolumn{2}{c}{ipog} & \multicolumn{2}{c|}{bot} & \multicolumn{2}{c}{ACTS} & \multicolumn{2}{c}{ipog} & \multicolumn{2}{c|}{bot} & \multicolumn{2}{c}{ACTS} & \multicolumn{2}{c}{ipog} & \multicolumn{2}{c}{pbot1G} \\
{} &    size & time &    size & time &      size & time &      size & time &      size & time &      size & time & size & time & size & time & size & time \\
\midrule
\cite{cohen2008constructing} \\
\midrule
1        &   48 &       9 &   45.00 &      7.06 &   46.00 &      7.45 &    293 &      12 &   289.40 &      8.59 &    289.80 &     68.41 &    1680 &    1207 &  1671.60 &    163.59 &   1662.40 &    843.20 \\
2        &   32 &       9 &   34.20 &      7.05 &   32.80 &      7.37 &    174 &       6 &   179.60 &      6.63 &    164.80 &     48.46 &     873 &     586 &   869.60 &     78.77 &    770.60 &    527.47 \\
3        &   19 &       9 &   19.80 &      4.32 &   18.80 &      4.36 &     71 &       2 &    66.60 &      3.80 &     67.40 &      4.45 &     212 &       2 &   211.20 &      1.97 &    202.80 &     25.36 \\
4        &   22 &       7 &   22.60 &      6.56 &   24.60 &      6.97 &    102 &       3 &   101.40 &      4.80 &    102.20 &     11.79 &     374 &      12 &   383.00 &     15.75 &    379.80 &    247.18 \\
\bottomrule
\end{tabular}
}
\caption{Comparison of IPOG in \emph{ACTS} and IPOG and BOT-its in \emph{CTLog} for strengths $t=\{2,3,4\}$}
\label{tab:acts-vs-ctlog}
\end{table*}

\begin{table*}
\centering
\resizebox{\textwidth}{!}{
\begin{tabular}{l|c|rr|rr|rr|rr|rr|rr}
\toprule
{} & {} & \multicolumn{4}{c|}{$t=2$} & \multicolumn{4}{c|}{$t=3$} & \multicolumn{4}{c}{$t=4$} \\
\midrule
{} & loading & \multicolumn{2}{c|}{ipog} & \multicolumn{2}{c|}{bot} & \multicolumn{2}{c|}{ipog} & \multicolumn{2}{c|}{bot} & \multicolumn{2}{c|}{ipog} & \multicolumn{2}{c}{bot} \\
{} & time &  size & time &   size & time &   size & time &    size & time &    size & time &    size & time \\
\midrule
unconstr           &         1.21                       &  \textbf{9.20} &            2.04 &          10.80 &   \textbf{1.90} &  \textbf{17.60} &   \textbf{1.82} &  \textbf{17.60} &             1.82 &  \textbf{44.40} &    \textbf{1.23} &           46.60 &             1.26 \\
\midrule
AProVE09-01     &        44.33  &  \textbf{9.20} &           46.60 &          10.80 &  \textbf{45.98} &  \textbf{17.60} &           46.23 &  \textbf{17.60} &   \textbf{45.09} &  \textbf{44.40} &            47.08 &           46.60 &   \textbf{46.50} \\
AProVE09-05     &        18.65  &  \textbf{4.00} &           22.17 &           4.60 &  \textbf{21.68} &   \textbf{4.00} &  \textbf{21.61} &            6.60 &            21.90 &   \textbf{4.00} &            21.36 &            6.60 &   \textbf{21.33} \\
AProVE09-07     &        10.81  &  \textbf{2.00} &           18.57 &           3.40 &  \textbf{18.26} &   \textbf{2.00} &           17.67 &            3.00 &   \textbf{16.90} &   \textbf{2.00} &            17.54 &            3.40 &   \textbf{16.84} \\
AProVE09-20     &        35.62  &  \textbf{9.20} &           89.50 &  \textbf{9.20} &  \textbf{79.86} &  \textbf{20.40} &          129.99 &           24.60 &  \textbf{126.91} &  \textbf{50.40} &  \textbf{203.39} &           63.20 &           204.72 \\
\bottomrule
\end{tabular}
}
\caption{Average test suite size and runtime for the instances with 10 parameters and 5000 conflicts}
\label{tab:10p-5k}
\bigskip
\resizebox{\textwidth}{!}{
\begin{tabular}{l|c|rr|rr|rr|rr|rr|rr}
\toprule
{} & {} & \multicolumn{4}{c|}{$t=2$} & \multicolumn{4}{c|}{$t=3$} & \multicolumn{4}{c}{$t=4$} \\
\midrule
{} & loading & \multicolumn{2}{c|}{ipog} & \multicolumn{2}{c|}{bot} & \multicolumn{2}{c|}{ipog} & \multicolumn{2}{c|}{bot} & \multicolumn{2}{c|}{ipog} & \multicolumn{2}{c}{bot} \\
{} & time & size & time &   size & time &   size & time &    size & time &    size & time &    size & time \\
\midrule
unconstr    &   1.21                            &   \textbf{9.20} &            2.04 &           10.80 &    \textbf{1.90} &  \textbf{17.60} &    \textbf{1.82} &  \textbf{17.60} &             1.82 &  \textbf{44.40} &    \textbf{1.23} &           46.60 &             1.26 \\
\midrule
AProVE09-08      & 12.53 &   \textbf{2.00} &  \textbf{23.29} &            4.00 &            26.52 &   \textbf{2.00} &   \textbf{22.52} &            4.00 &            25.39 &   \textbf{2.00} &   \textbf{22.87} &            3.80 &            25.80 \\
AProVE09-15      & 88.61 &   \textbf{7.60} &          106.83 &            9.20 &  \textbf{102.89} &  \textbf{16.60} &           114.97 &           19.20 &  \textbf{104.82} &  \textbf{35.00} &           123.00 &           39.00 &  \textbf{112.11} \\
AProVE09-20      & 36.53 &   \textbf{8.40} &           73.80 &           10.00 &   \textbf{71.01} &  \textbf{18.20} &            83.99 &           24.00 &   \textbf{82.64} &  \textbf{42.20} &           104.05 &           53.40 &  \textbf{102.23} \\
AProVE09-21      & 34.06 &  \textbf{10.40} &           90.60 &           11.60 &   \textbf{84.96} &  \textbf{22.60} &  \textbf{113.97} &           28.80 &           119.04 &  \textbf{42.40} &           155.57 &           51.80 &  \textbf{117.91} \\
\bottomrule
\end{tabular}
}
\caption{Average test suite size and runtime for the instances with 10 parameters and 10000 conflicts}
\label{tab:10p-10k}
\end{table*}

\begin{table*}
\centering
\resizebox{\textwidth}{!}{
\begin{tabular}{l|c|rr|rr|rr|rr|rr|rr}
\toprule
{} & {} & \multicolumn{4}{c|}{$t=2$} & \multicolumn{4}{c|}{$t=3$} & \multicolumn{4}{c}{$t=4$} \\
\midrule
{} & loading & \multicolumn{2}{c|}{ipog} & \multicolumn{2}{c|}{bot} & \multicolumn{2}{c|}{ipog} & \multicolumn{2}{c|}{bot} & \multicolumn{2}{c|}{ipog} & \multicolumn{2}{c}{bot} \\
{} & time & size & time &   size & time &   size & time &    size & time &    size & time &    size & time \\
\midrule
unconstr   &         1.15                       &  \textbf{10.40} &             2.04 &  15.20 &   \textbf{1.92} &           27.60 &    \textbf{1.92} &  \textbf{26.00} &            2.03 &   \textbf{66.60} &    \textbf{1.24} &            68.40 &             3.55 \\
\midrule
AProVE09-01      & 44.05 &  \textbf{10.40} &   \textbf{45.28} &  15.80 &           45.69 &           27.40 &            48.86 &  \textbf{25.60} &  \textbf{46.07} &            64.40 &            52.15 &   \textbf{64.00} &   \textbf{51.17} \\
AProVE09-05      & 18.50 &   \textbf{5.20} &            22.49 &   5.40 &  \textbf{21.92} &   \textbf{7.00} &            22.34 &            8.20 &  \textbf{21.70} &    \textbf{7.00} &            25.79 &             9.60 &   \textbf{22.64} \\
AProVE09-07      & 10.17 &   \textbf{5.00} &   \textbf{21.51} &   8.00 &           23.66 &   \textbf{5.00} &   \textbf{22.52} &            9.60 &           22.97 &    \textbf{5.00} &            24.62 &             9.20 &   \textbf{24.41} \\
AProVE09-20      & 35.50 &  \textbf{13.20} &  \textbf{149.72} &  14.60 &          198.81 &  \textbf{44.00} &  \textbf{240.58} &           62.20 &          304.47 &  \textbf{144.00} &  \textbf{383.80} &           203.80 &           530.91 \\
\bottomrule
\end{tabular}
}
\caption{Average test suite size and runtime for the instances with 20 parameters and 5000 conflicts}
\label{tab:20p-5k}
\bigskip
\resizebox{\textwidth}{!}{
\begin{tabular}{l|c|rr|rr|rr|rr|rr|rr}
\toprule
{} & {} & \multicolumn{4}{c|}{$t=2$} & \multicolumn{4}{c|}{$t=3$} & \multicolumn{4}{c}{$t=4$} \\
\midrule
{} & loading & \multicolumn{2}{c|}{ipog} & \multicolumn{2}{c|}{bot} & \multicolumn{2}{c|}{ipog} & \multicolumn{2}{c|}{bot} & \multicolumn{2}{c|}{ipog} & \multicolumn{2}{c}{bot} \\
{} & time & size & time &   size & time &   size & time &    size & time &    size & time &    size & time \\
\midrule
unconstr   & 1.15                               &  \textbf{10.40} &             2.04 &  15.20 &    \textbf{1.92} &           27.60 &    \textbf{1.92} &  \textbf{26.00} &             2.03 &   \textbf{66.60} &    \textbf{1.24} &            68.40 &             3.55 \\
\midrule
AProVE09-08      & 12.84 &   \textbf{4.00} &   \textbf{25.24} &   5.60 &            26.37 &   \textbf{4.00} &   \textbf{24.21} &            6.20 &            25.20 &    \textbf{4.00} &            27.32 &             6.40 &   \textbf{26.50} \\
AProVE09-15      & 89.03 &  \textbf{11.00} &           135.67 &  12.40 &  \textbf{118.28} &  \textbf{30.40} &           187.69 &           35.60 &  \textbf{151.98} &   \textbf{75.60} &           302.64 &            92.60 &  \textbf{210.73} \\
AProVE09-20      & 36.24 &  \textbf{14.80} &   \textbf{91.63} &  16.80 &            97.64 &  \textbf{45.60} &  \textbf{118.92} &           56.00 &           125.49 &  \textbf{133.00} &           174.81 &           151.40 &  \textbf{156.34} \\
AProVE09-21      & 32.79 &  \textbf{15.60} &  \textbf{148.43} &  20.00 &           187.56 &  \textbf{51.40} &  \textbf{263.75} &           65.20 &           267.28 &  \textbf{129.80} &           350.76 &           148.00 &  \textbf{313.68} \\
\bottomrule
\end{tabular}
}
\caption{Average test suite size and runtime for the instances with 20 parameters and 10000 conflicts}
\label{tab:20p-10k}
\end{table*}

\begin{table*}
\centering
\resizebox{\textwidth}{!}{
\begin{tabular}{l|rr|rr|rr|rr|rr|rr}
\toprule
{} & \multicolumn{4}{c|}{$t=2$} & \multicolumn{4}{c|}{$t=3$} & \multicolumn{4}{c}{$t=4$} \\
\midrule
{} & \multicolumn{2}{c|}{ipog} & \multicolumn{2}{c|}{bot} & \multicolumn{2}{c|}{ipog} & \multicolumn{2}{c|}{bot} & \multicolumn{2}{c|}{ipog} & \multicolumn{2}{c}{bot} \\
{} &  size & time &   size & time &   size & time &    size & time &    size & time &    size & time \\
\midrule
unconstr                                  &  \textbf{12.00} &    \textbf{2.06} &           16.60 &             2.07 &           30.20 &    \textbf{1.96} &  \textbf{29.00} &             2.16 &   \textbf{73.80} &    \textbf{1.41} &           78.20 &             8.20 \\
\midrule
AProVE09-01      &  \textbf{11.80} &   \textbf{47.73} &           16.40 &            49.51 &           29.80 &            48.88 &  \textbf{28.40} &   \textbf{46.12} &   \textbf{72.20} &   \textbf{57.81} &           72.40 &            58.27 \\
AProVE09-05      &   \textbf{8.60} &            23.65 &            8.80 &   \textbf{23.51} &  \textbf{11.80} &            22.81 &           13.20 &   \textbf{21.67} &   \textbf{17.00} &            30.22 &           20.00 &   \textbf{26.88} \\
AProVE09-07      &   \textbf{9.00} &   \textbf{23.29} &           10.00 &            25.93 &  \textbf{11.00} &   \textbf{22.29} &           16.00 &            23.97 &   \textbf{11.00} &   \textbf{28.11} &           17.40 &            28.21 \\
AProVE09-20      &  \textbf{14.20} &  \textbf{163.77} &           16.40 &           208.08 &  \textbf{56.60} &  \textbf{265.15} &           76.80 &           309.68 &  \textbf{205.20} &  \textbf{429.67} &          258.00 &           432.94 \\
\bottomrule
\end{tabular}
}
\caption{Average test suite size and runtime for the instances with 25 parameters and 5000 conflicts}
\label{tab:25p-5k}
\bigskip
\resizebox{\textwidth}{!}{
\begin{tabular}{l|rr|rr|rr|rr|rr|rr}
\toprule
{} & \multicolumn{4}{c|}{$t=2$} & \multicolumn{4}{c|}{$t=3$} & \multicolumn{4}{c}{$t=4$} \\
\midrule
{} & \multicolumn{2}{c|}{ipog} & \multicolumn{2}{c|}{bot} & \multicolumn{2}{c|}{ipog} & \multicolumn{2}{c|}{bot} & \multicolumn{2}{c|}{ipog} & \multicolumn{2}{c}{bot} \\
{} &  size & time &   size & time &   size & time &    size & time &    size & time &    size & time \\
\midrule
unconstr                                  &  \textbf{12.00} &    \textbf{2.06} &          16.60 &             2.07 &           30.20 &    \textbf{1.96} &  \textbf{29.00} &             2.16 &   \textbf{73.80} &    \textbf{1.41} &           78.20 &             8.20 \\
\midrule
AProVE09-08     &   \textbf{5.00} &   \textbf{23.38} &           5.40 &            23.64 &   \textbf{6.00} &   \textbf{23.93} &            8.00 &            24.52 &    \textbf{6.00} &            32.07 &            8.00 &   \textbf{25.42} \\
AProVE09-15     &  \textbf{14.20} &           150.26 &          16.80 &  \textbf{127.13} &  \textbf{51.20} &           325.89 &           56.80 &  \textbf{198.42} &  \textbf{147.20} &           879.23 &          165.20 &  \textbf{339.15} \\
AProVE09-20     &  \textbf{16.80} &   \textbf{99.77} &          19.40 &           103.42 &  \textbf{62.80} &  \textbf{125.89} &           72.20 &           128.04 &  \textbf{194.00} &           224.04 &          201.00 &  \textbf{188.70} \\
AProVE09-21     &  \textbf{18.40} &  \textbf{178.47} &          22.80 &           203.77 &  \textbf{63.20} &  \textbf{289.73} &           79.20 &           361.22 &  \textbf{173.00} &  \textbf{457.32} &          207.20 &           457.79 \\
\bottomrule
\end{tabular}
}
\caption{Average test suite size and runtime for the instances with 25 parameters and 10000 conflicts}
\label{tab:25p-10k}
\end{table*}

\begin{table*}
\centering
\resizebox{\textwidth}{!}{
\begin{tabular}{l|rr|rr|rr|rr|rr|rr}
\toprule
{} & \multicolumn{4}{c|}{$t=2$} & \multicolumn{4}{c|}{$t=3$} & \multicolumn{4}{c}{$t=4$} \\
\midrule
{} & \multicolumn{2}{c|}{ipog} & \multicolumn{2}{c|}{bot} & \multicolumn{2}{c|}{ipog} & \multicolumn{2}{c|}{bot} & \multicolumn{2}{c|}{ipog} & \multicolumn{2}{c}{bot} \\
{} &  size & time &   size & time &   size & time &    size & time &    size & time &    size & time \\
\midrule
unconstr                                  &  \textbf{14.00} &             2.14 &           22.80 &   \textbf{2.05} &            39.40 &    \textbf{1.92} &   \textbf{38.80} &             4.99 &  \textbf{101.00} &    \textbf{1.87} &           107.80 &            156.74 \\
\midrule
AProVE09-01      &  \textbf{13.80} &            50.14 &           22.80 &  \textbf{49.20} &            38.60 &            61.01 &   \textbf{37.60} &   \textbf{51.89} &   \textbf{98.40} &   \textbf{92.27} &            99.40 &            210.19 \\
AProVE09-05      &  \textbf{14.80} &   \textbf{24.90} &           16.20 &           25.69 &   \textbf{34.20} &   \textbf{29.91} &            38.40 &            31.05 &   \textbf{55.80} &  \textbf{172.02} &            62.00 &            191.55 \\
AProVE09-07      &  \textbf{12.00} &   \textbf{22.19} &           14.20 &           24.73 &   \textbf{19.60} &            29.70 &            25.20 &   \textbf{28.42} &   \textbf{24.20} &           152.07 &            31.80 &   \textbf{106.58} \\
AProVE09-20      &  \textbf{30.60} &  \textbf{163.00} &           34.00 &          185.96 &  \textbf{154.40} &           386.53 &           162.40 &  \textbf{281.93} &           677.80 &          4592.67 &  \textbf{642.80} &  \textbf{1128.42} \\
\bottomrule
\end{tabular}
}
\caption{Average test suite size and runtime for the instances with 50 parameters and 5000 conflicts}
\label{tab:50p-5k}
\bigskip
\resizebox{\textwidth}{!}{
\begin{tabular}{l|rr|rr|rr|rr|rr|rr}
\toprule
{} & \multicolumn{4}{c|}{$t=2$} & \multicolumn{4}{c|}{$t=3$} & \multicolumn{4}{c}{$t=4$} \\
\midrule
{} & \multicolumn{2}{c|}{ipog} & \multicolumn{2}{c|}{bot} & \multicolumn{2}{c|}{ipog} & \multicolumn{2}{c|}{bot} & \multicolumn{2}{c|}{ipog} & \multicolumn{2}{c}{bot} \\
{} &  size & time &   size & time &   size & time &    size & time &    size & time &    size & time \\
\midrule
unconstr                                  &  \textbf{14.00} &             2.14 &          22.80 &    \textbf{2.05} &            39.40 &    \textbf{1.92} &   \textbf{38.80} &             4.99 &   \textbf{101.00} &    \textbf{1.87} &           107.80 &            156.74 \\
\midrule
AProVE09-08     &   \textbf{6.00} &   \textbf{25.01} &  \textbf{6.00} &            25.92 &    \textbf{8.00} &            28.71 &             9.40 &   \textbf{25.66} &     \textbf{8.00} &           145.87 &             9.40 &    \textbf{80.84} \\
AProVE09-15     &  \textbf{36.40} &           347.93 &          43.60 &  \textbf{235.99} &  \textbf{181.20} &          1746.70 &           202.00 &  \textbf{760.56} &   \textbf{709.75} &         20900.35 &           742.00 &  \textbf{2452.07} \\
AProVE09-20     &  \textbf{31.40} &  \textbf{128.28} &          42.60 &           163.22 &  \textbf{164.60} &           355.35 &           177.60 &  \textbf{225.90} &            658.20 &          4026.96 &  \textbf{641.80} &  \textbf{1237.86} \\
AProVE09-21     &  \textbf{37.60} &  \textbf{274.70} &          41.40 &           321.27 &  \textbf{165.40} &           819.80 &           195.00 &  \textbf{733.68} &   \textbf{661.80} &          6802.09 &           712.60 &  \textbf{2176.96} \\
\bottomrule
\end{tabular}
}
\caption{Average test suite size and runtime for the instances with 50 parameters and 10000 conflicts}
\label{tab:50p-10k}
\end{table*}

\begin{table*}
\centering
\resizebox{\textwidth}{!}{
\begin{tabular}{l|rr|rr|rr|rr|rr|rr}
\toprule
{} & \multicolumn{4}{c|}{$t=2$} & \multicolumn{4}{c|}{$t=3$} & \multicolumn{4}{c}{$t=4$} \\
\midrule
{} & \multicolumn{2}{c|}{ipog} & \multicolumn{2}{c|}{bot} & \multicolumn{2}{c|}{ipog} & \multicolumn{2}{c|}{bot} & \multicolumn{2}{c|}{ipog} & \multicolumn{2}{c}{pbot1G} \\
{} &  size & time &   size & time &   size & time &    size & time &    size & time &    size & time \\
\midrule
unconstr                                 &  \textbf{16.00} &    \textbf{2.43} &           32.00 &             2.58 &   \textbf{48.40} &   \textbf{2.37} &            49.40 &             33.32 &  \textbf{131.60} &   \textbf{17.02} &            146.00 &            430.86 \\
\midrule
AProVE09-01     &  \textbf{16.00} &            61.29 &           26.60 &   \textbf{54.45} &            47.40 &  \textbf{89.57} &   \textbf{46.80} &             91.63 &  \textbf{129.00} &  \textbf{280.06} &            134.60 &            564.85 \\
AProVE09-05     &  \textbf{27.40} &   \textbf{31.82} &           28.80 &            32.47 &   \textbf{71.60} &          119.94 &            77.80 &    \textbf{89.96} &  \textbf{117.80} &          2897.94 &            134.00 &  \textbf{1204.98} \\
AProVE09-07     &  \textbf{19.80} &   \textbf{24.43} &           22.00 &            25.12 &   \textbf{43.40} &           65.01 &            50.40 &    \textbf{55.23} &   \textbf{60.80} &          2579.34 &             70.40 &  \textbf{1166.53} \\
AProVE09-20     &  \textbf{54.80} &  \textbf{212.28} &           64.20 &           219.54 &           396.40 &         3585.23 &  \textbf{370.40} &   \textbf{726.71} &                - &                - &  \textbf{2141.80} &  \textbf{7813.83} \\
\bottomrule
\end{tabular}
}
\caption{Average test suite size and runtime for the instances with 100 parameters and 5000 conflicts}
\label{tab:100p-5k}
\bigskip
\resizebox{\textwidth}{!}{
\begin{tabular}{l|rr|rr|rr|rr|rr|rr}
\toprule
{} & \multicolumn{4}{c|}{$t=2$} & \multicolumn{4}{c|}{$t=3$} & \multicolumn{4}{c}{$t=4$} \\
\midrule
{} & \multicolumn{2}{c|}{ipog} & \multicolumn{2}{c|}{bot} & \multicolumn{2}{c|}{ipog} & \multicolumn{2}{c|}{bot} & \multicolumn{2}{c|}{ipog} & \multicolumn{2}{c}{pbot1G} \\
{} &  size & time &   size & time &   size & time &    size & time &    size & time &    size & time \\
\midrule
unconstr                                 &  \textbf{16.00} &    \textbf{2.43} &           32.00 &             2.58 &   \textbf{48.40} &  \textbf{2.37} &            49.40 &             33.32 &  \textbf{131.60} &  \textbf{17.02} &            146.00 &             430.86 \\
\midrule
AProVE09-08    &   \textbf{7.00} &   \textbf{24.26} &            7.20 &            27.60 &    \textbf{8.00} &          51.17 &            10.00 &    \textbf{49.14} &    \textbf{8.00} &         1583.95 &             10.80 &    \textbf{968.71} \\
AProVE09-15    &  \textbf{78.20} &           845.51 &           91.00 &  \textbf{538.64} &  \textbf{540.80} &       16280.83 &           595.80 &  \textbf{3076.52} &                - &               - &  \textbf{2992.00} &  \textbf{21327.25} \\
AProVE09-20    &  \textbf{56.40} &           187.74 &           63.20 &  \textbf{168.65} &           390.20 &        2627.94 &  \textbf{372.20} &   \textbf{514.98} &                - &               - &  \textbf{1907.40} &   \textbf{5553.26} \\
AProVE09-21    &  \textbf{78.20} &  \textbf{410.39} &           93.80 &           571.26 &  \textbf{459.00} &        4887.11 &           491.20 &  \textbf{1350.52} &                - &               - &  \textbf{2130.20} &   \textbf{7857.80} \\
\bottomrule
\end{tabular}
}
\caption{Average test suite size and runtime for the instances with 100 parameters and 10000 conflicts}
\label{tab:100p-10k}
\end{table*}

We set the constants $\Delta_\mathcal{A}$ and $\nabla_\mathcal{A}$ to 10 and 5 respectively, $\mathcal{S}$ to five random seeds and $MAX\_TRIES$ to 100. Additionally, we limited each query to the SAT solver to 300s, so if one of these queries cannot be completed we consider the formula as UNSAT.

We generated benchmarks with 10, 20, 25, 50 and 100 parameters of domain 2 using a limit $c_{max}$ of 5000 and 10000 conflicts, setting $c_{min}$ always to $c_{max}/2$ (see Section~\ref{sec:gen-bench-sat} for more details regarding the generation parameters). In total, we generated 140 crafted benchmarks that we will make available to the community, as well as the SUT-G generator. 

Our experimentation environment consists of a cluster of nodes with two AMD 7403 processors each (24 cores at 2.8GHz) and 21 GB of RAM per core. Tables~\ref{tab:25p-5k},~\ref{tab:25p-10k},~\ref{tab:50p-5k},~\ref{tab:50p-10k},~\ref{tab:100p-5k} and~\ref{tab:100p-10k} show the results of our experimentation. We show in bold the best results in terms of average test suite size and time. We run each benchmark with 5 different seeds to mitigate the stochastic behaviour of the tested algorithms, using a time limit of 8h and a memory limit of 21 GB. Additionally, we executed each benchmark for strengths $t=\{2,3,4\}$.

In the following sections, we will analyze the results of our experimental evaluation.

\subsection{Comparing SUT-G with random SUT generators}\label{sec:rand-gen-exp}

Recently, the CT community has organized the first edition of the CT Competition\footnote{CT-Competition: \url{https://fmselab.github.io/ct-competition/}}, where different CT tools are evaluated over different tracks (i.e. sets of SUT models with different properties).
In this first edition, a random SUT generator was used to provide the benchmarks for these tracks\footnote{The random benchmark generator is available here: \url{https://github.com/fmselab/CIT_Benchmark_Generator}}.

We will compare SUT-G with the benchmarks of the \emph{BOOLC} track of the CT Competition 2022, which consists of SUTs that only contain boolean parameters and a set of random SUT constraints.
Notice that the SUTs generated by SUT-G could be included in this track.

We noticed that the maximum number of parameters that we can find in the SUTs of \emph{BOOLC} is 20.
We discarded those benchmarks where the number of parameters was less than 10, and generated SUTs with 10 and 20 parameters using SUT-G to try to perform a fair comparison.

We ran the IPOG and BOT-its algorithms over all these benchmarks to analyze their behaviour.
Table~\ref{tab:boolc} shows the results of the \emph{BOOLC} track, and tables~\ref{tab:10p-5k}, \ref{tab:10p-10k}, \ref{tab:20p-5k}, \ref{tab:20p-10k} show the results of the SUT-G benchmarks for 10 and 20 parameters using a conflicts limit of 5000 and 10000 conflicts in both cases.

\begin{table*}[ht]
\centering
\begin{tabular}{l|rr|rr|rr|rr|rr|rr}
\toprule
{} & \multicolumn{4}{c|}{$t=2$} & \multicolumn{4}{c|}{$t=3$} & \multicolumn{4}{c}{$t=4$} \\
\midrule
{} & \multicolumn{2}{c|}{ipog} & \multicolumn{2}{c|}{bot} & \multicolumn{2}{c|}{ipog} & \multicolumn{2}{c|}{bot} & \multicolumn{2}{c|}{ipog} & \multicolumn{2}{c}{bot} \\
{} &   size & time &   size & time &   size & time &   size & time &    size & time &    size & time \\
\midrule
0  &   9.2 &      1.29 &   8.8 &      1.44 &  22.2 &      1.59 &  22.0 &      1.53 &   37.2 &      1.76 &   37.6 &      1.84 \\
1  &   9.6 &      2.24 &   9.6 &      1.63 &  22.4 &      2.40 &  20.8 &      2.47 &   52.8 &      1.87 &   51.0 &      2.21 \\
2  &   8.8 &      2.20 &   8.8 &      1.68 &  19.0 &      2.27 &  18.8 &      2.03 &   44.8 &      2.26 &   42.8 &      2.65 \\
3  &   8.4 &      2.24 &   8.4 &      1.57 &  20.8 &      2.23 &  20.0 &      2.41 &   43.8 &      2.14 &   42.2 &      2.19 \\
4  &  10.2 &      1.29 &  10.2 &      1.42 &  31.2 &      1.69 &  29.4 &      1.86 &   85.4 &      1.92 &   80.8 &      4.03 \\
5  &   1.0 &      2.21 &   1.0 &      1.99 &   1.0 &      1.54 &   1.0 &      1.55 &    1.0 &      1.81 &    1.0 &      1.85 \\
6  &  13.6 &      2.24 &  12.8 &      1.58 &  30.4 &      2.61 &  30.2 &      2.30 &   48.6 &      2.26 &   49.8 &      2.55 \\
7  &   9.6 &      1.78 &   9.8 &      1.65 &  25.0 &      2.39 &  24.2 &      2.14 &   59.6 &      2.27 &   59.8 &      2.72 \\
8  &   8.2 &      2.25 &   8.4 &      1.73 &  17.4 &      2.30 &  16.6 &      2.35 &   36.0 &      1.49 &   36.8 &      1.55 \\
9  &  13.0 &      2.27 &  12.6 &      1.60 &  36.8 &      2.35 &  35.8 &      2.56 &  101.8 &      1.60 &   95.0 &      3.10 \\
10 &   8.0 &      2.19 &   7.8 &      1.83 &  18.6 &      2.20 &  18.6 &      2.39 &   37.8 &      2.00 &   37.6 &      1.94 \\
12 &   4.0 &      1.31 &   4.0 &      1.44 &   4.0 &      1.55 &   4.0 &      1.69 &    4.0 &      1.81 &    4.0 &      1.78 \\
14 &   7.4 &      2.19 &   7.0 &      1.53 &  13.4 &      2.35 &  14.6 &      2.56 &   26.4 &      2.11 &   27.4 &      1.95 \\
15 &  10.8 &      2.13 &  11.0 &      1.57 &  32.8 &      2.15 &  31.4 &      2.35 &   80.2 &      2.07 &   79.2 &      2.65 \\
16 &   7.0 &      2.03 &   6.8 &      2.05 &   9.0 &      1.41 &   9.0 &      1.52 &    9.0 &      2.27 &    9.0 &      2.32 \\
17 &   9.8 &      1.18 &  10.4 &      1.60 &  24.6 &      1.72 &  23.6 &      1.71 &   58.2 &      1.68 &   58.4 &      1.96 \\
18 &   5.0 &      2.05 &   5.0 &      1.86 &   6.0 &      1.43 &   6.0 &      1.44 &    6.0 &      1.76 &    6.0 &      1.83 \\
19 &  14.0 &      2.28 &  13.8 &      1.80 &  40.6 &      2.44 &  41.6 &      2.52 &  101.8 &      2.43 &  103.8 &      4.01 \\
20 &  10.2 &      2.24 &  11.2 &      1.68 &  28.0 &      2.17 &  26.8 &      2.17 &   75.6 &      2.28 &   70.0 &      3.68 \\
21 &   5.6 &      2.25 &   5.2 &      1.72 &   8.0 &      2.23 &   8.0 &      2.34 &    8.0 &      1.96 &    8.0 &      2.06 \\
22 &  11.0 &      2.16 &  12.4 &      1.69 &  30.2 &      2.38 &  29.8 &      2.54 &   81.8 &      2.25 &   76.8 &      3.85 \\
25 &   6.0 &      1.27 &   5.8 &      1.30 &  10.0 &      1.55 &  10.0 &      1.63 &   12.0 &      1.92 &   12.0 &      1.79 \\
26 &   6.0 &      1.84 &   6.0 &      1.72 &  10.4 &      1.19 &  11.4 &      1.21 &   16.0 &      1.66 &   16.0 &      2.01 \\
28 &   9.8 &      2.19 &  10.0 &      1.61 &  24.8 &      2.18 &  25.2 &      2.21 &   64.2 &      2.01 &   61.6 &      2.10 \\
30 &  11.6 &      1.83 &  11.0 &      1.70 &  37.6 &      2.68 &  35.6 &      2.82 &  108.4 &      2.40 &  103.6 &      4.04 \\
31 &   6.6 &      2.24 &   6.2 &      1.60 &  10.8 &      2.41 &  10.2 &      2.47 &   12.0 &      2.32 &   12.0 &      2.43 \\
33 &  11.0 &      2.32 &  10.2 &      1.72 &  31.2 &      2.55 &  29.2 &      2.57 &   86.2 &      2.23 &   78.4 &      3.39 \\
34 &  10.0 &      2.30 &  10.2 &      1.69 &  27.8 &      2.30 &  27.8 &      2.18 &   77.2 &      1.81 &   74.0 &      2.76 \\
36 &   9.4 &      2.25 &  10.0 &      1.54 &  26.8 &      2.39 &  25.6 &      2.43 &   71.0 &      1.74 &   69.4 &      2.25 \\
38 &   8.8 &      1.48 &   8.4 &      1.95 &  21.6 &      1.38 &  22.0 &      1.45 &   52.0 &      1.44 &   50.8 &      1.58 \\
40 &  10.2 &      2.28 &  10.2 &      1.73 &  30.0 &      1.90 &  28.6 &      1.88 &   81.4 &      2.27 &   77.2 &      3.19 \\
41 &   5.6 &      2.18 &   5.6 &      1.59 &   8.0 &      2.26 &   8.0 &      2.27 &    8.0 &      2.46 &    8.0 &      2.93 \\
42 &   6.0 &      1.29 &   6.0 &      1.65 &   6.0 &      1.68 &   6.0 &      1.70 &    6.0 &      1.79 &    6.0 &      1.82 \\
43 &  11.4 &      2.36 &  10.4 &      1.84 &  32.4 &      1.50 &  32.2 &      1.60 &   90.4 &      2.08 &   88.8 &      3.32 \\
44 &   9.0 &      2.24 &   9.2 &      1.86 &  20.8 &      1.39 &  20.4 &      1.68 &   49.0 &      2.22 &   47.8 &      2.91 \\
45 &  10.0 &      1.33 &  10.6 &      1.70 &  25.4 &      1.85 &  25.2 &      1.91 &   65.0 &      2.06 &   62.2 &      3.41 \\
47 &  13.0 &      2.33 &  12.2 &      1.73 &  40.4 &      2.17 &  38.0 &      2.23 &  111.2 &      2.40 &  104.2 &      4.34 \\
48 &   9.4 &      2.24 &  10.0 &      1.68 &  24.8 &      2.17 &  25.4 &      2.28 &   62.2 &      1.92 &   63.8 &      2.30 \\
\bottomrule
\end{tabular}
\caption{Results for the IPOG and BOT-its algorithms over the \emph{BOOLC} track of the CT Competition 2022. Benchmarks with less than 10 parameters were discarded.}
\label{tab:boolc}
\end{table*}

We observe that all the algorithms are able to report a test suite after at most 4 seconds for all the strengths in all the SUTs in the \emph{BOOLC} track.
This indirectly shows that the generated SUT constraints for this track are not challenging enough for the tested algorithms.
On the other hand, we do observe variability in the generated test suite sizes, which means that SUT constraints are producing forbidden tuples that modify their structure.

Regarding SUT-G, the first thing that we notice is much higher time consumption.
To discard a possible bias produced by the loading times of the SUT models we included an additional column with these values in tables~\ref{tab:10p-5k}, \ref{tab:10p-10k}, \ref{tab:20p-5k}, \ref{tab:20p-10k}.
We observe much higher variability in the run times with respect to \emph{BOOLC}, even after discounting the loading times.
This shows that the constraints generated with SUT-G produce a higher impact on the tested MCAC algorithms.

Additionally, we observe a significant increase of the test suite size in certain benchmarks (especially \emph{AProVE09-20} and \emph{AProVE09-21} in tables~\ref{tab:20p-5k} and~\ref{tab:20p-10k}, where the test suite size is doubled compared to the unconstrained versions for strengths $t>2$).

\subsection{Impact of the SUT constraints}
\label{sec:impact-sut-constr}

The second question we address is how these SUT constraints impact the performance of the tested algorithms. To analyze this, we included an execution of a benchmark with the same number of parameters but without SUT constraints on the \emph{unconstr} row of Tables~\ref{tab:10p-5k}, \ref{tab:10p-10k}, \ref{tab:20p-5k}, \ref{tab:20p-10k},  \ref{tab:25p-5k}, \ref{tab:25p-10k}, \ref{tab:50p-5k}, \ref{tab:50p-10k}, \ref{tab:100p-5k} and~\ref{tab:100p-10k}.

In general, we observe that the addition of SUT constraints has a negative impact on the generation runtime of the MCAC, regardless of the number of parameters of the SUT, the strength $t$ or the MCAC algorithm. This is expected, as at some point we need to ensure that the test case that we are building is consistent with the SUT constraints. This negative effect is amplified when we increase the number of parameters, strength or \emph{hardness} of the SUT constraints (i.e. number of conflicts).

However, we found some cases where the run time is reduced, especially in some executions for $t=4$ on the BOT-its and PBOT-its algorithms (for example, benchmark \emph{AProVE09-08} in Table~\ref{tab:50p-10k} for $t=4$ and BOT-its, or benchmark \emph{q\_query\_3\_l43} in Table~\ref{tab:50p-5k} for $t=4$ and BOT-its).
This suggests that some of the efficiencies on the BOT-its algorithms regarding SUT constraints handling might benefit its performance in some circumstances.

Regarding the test suite size, we observe more variability than with run time.
In some cases, it increases (for example benchmark \emph{AProVE09-21} in Table~\ref{tab:50p-5k} for all the strengths and algorithms), while in others it decreases (for example benchmark \emph{UR-10} in Table~\ref{tab:25p-10k} for all strengths and algorithms). In the first case, there might be some combinations that belong to some tests of the unconstrained version that are now forbidden, so these tests need to be \emph{split} into different tests, increasing the overall test suite size.
For the second case, the generated benchmark is so constrained that it is forbidding some of the possible tests that we can find in the unconstrained version, reducing the final test suite size.

Notice that the \emph{IPOG} and \emph{BOT-its} algorithms cannot guarantee the optimality of the reported test suite (i.e. they cannot certify the \emph{CAN}), so solutions with smaller test suites may exist. Nonetheless, these results provide interesting insights about the behaviour of practical MCAC generation algorithms when dealing with \emph{hard} SUT constraints, as complete approaches that can obtain optimal MCACs such as MaxSAT MCAC~\cite{ansotegui_incomplete_2022,AnsoteguiIMJ13} or the CALOT~\cite{YamadaKACOB15} algorithm will struggle with these kinds of benchmarks (see Section~\ref{sec:recipes-mcac} for more in-depth discussion about this).

\subsection{Comparing the performance of the IPOG and (P)BOT-its algorithms}
\label{sec:exps-ipog-bot}

In Section~\ref{sec:impact-sut-constr} we analyzed the overall impact of the generated benchmarks over their unconstrained version.
In this section, we analyze the differences in performance for the \emph{IPOG} and \emph{(P)BOT-its} algorithms (see Sections~\ref{sec:ipog} and~\ref{sec:bot} respectively).

In general, we observe how \emph{IPOG} obtains better sizes than \emph{BOT-its} (especially with strength $t<4$) but worse run times. As we discussed in previous sections, the \emph{BOT-its} algorithm heuristically creates one test at a time that might be inconsistent during its construction, and later \emph{amends} it to try to reduce the number of SAT solver queries. It is clear that this procedure reduces the efficiency of the heuristic and therefore can increase the final test suite size, as some of its choices might be suboptimal once the test case is repaired. On the other hand, its final run time is reduced, especially when dealing with \emph{hard} SUT constraints, as SAT solver queries are also reduced with respect to \emph{IPOG}.

This is an interesting result, as in previous works the \emph{IPOG} algorithm obtained better run times than \emph{BOT-its} on the original state-of-the-art benchmarks~\cite{Yamada16,DBLP:conf/cp/AnsoteguiOT21}.
We also observe this behaviour in the results of Table~\ref{tab:acts-vs-ctlog}, where we compared the two mentioned algorithms using 58 state-of-the-art SUT models.

In Tables~\ref{tab:100p-5k} and~\ref{tab:100p-10k}, where we are solving SUT models with 100 parameters, we observe the limitations in terms of memory consumption of both \emph{IPOG} and \emph{BOT-its}. When we increase the strength $t$ above 3, \emph{IPOG} is only able to report an MCAC for 15 of the 28 benchmarks, while the basic \emph{BOT-its} could not report any of them\footnote{We did not include these results in the tables. Instead, we just report the results of PBOT-its}. This is why we executed the \emph{pool} version of the \emph{BOT-its} algorithm with a pool of 1GB (referred to as \emph{pbot1G} in the mentioned tables, see~\cite{DBLP:conf/cp/AnsoteguiOT21}). In contrast, this algorithm can solve all benchmarks except 2.

In these cases, we observe larger differences in runtime between IPOG and PBOT-its, being of up to an order of magnitude for benchmarks \emph{q\_query\_3\_L60}, \emph{q\_query\_3\_l39} and \emph{q\_query\_3\_l40} in Table~\ref{tab:100p-5k}, and \emph{q\_query\_3\_l40} and \emph{q\_query\_3\_l43} in Table~\ref{tab:100p-10k} for $t=4$.

With the latest version of the PRBOT-its algorithm, we observe how it is able to outperform in most cases the test suite sizes of IPOG, especially for higher strengths ($t>3$) and higher number of parameters (50 or more).

Finally, we observe that there is a particular benchmark that is never solved (benchmark \emph{rbcl\_xits\_17}). Although we tried to adapt the hardness of the generated benchmarks in Algorithm~\ref{alg:sut-gen}, there might be cases where this is not always possible. In particular, it seems that in this benchmark the algorithms try to explore an exponential branch of the formula when completing certain test cases.
Notice that the \emph{IPOG} algorithm would not be able to avoid these branches.
On the other hand, \emph{BOT-its} could be modified to try to mitigate this effect by adapting the \emph{amend} procedure on each test. This can be considered future work.

Thanks to the development of the \emph{SUT-G} generator, we have been able to provide another point of view on the performance of two well-known MCAC generation algorithms.
As result, in the next section, we will try to provide some recipes to effectively apply MCAC algorithms in real-world scenarios.


\section{Recipes for Choosing  MCAC Tools}
\label{sec:recipes-mcac}

In this section, we provide several interesting insights about the application of MCAC generation tools.
We extracted them thanks to the experimental evaluation that we conducted using the \emph{SUT-G} generator instances (see Section~\ref{sec:gen-bench-sat}).
These instances provided another point of view on the behaviour of the tested algorithm with respect to the currently available CT benchmarks.
Notice that in this case, we focus on the MCAC problem, but other CT problems can also be considered for this analysis.

In~\cite{DBLP:conf/cp/AnsoteguiOT21,Yamada16}, authors show how IPOG is faster than the BOT-its/\emph{Algorithm 5} algorithm for the current CT benchmarks. However, as we discussed in Section~\ref{sec:exps-ipog-bot}, we observe now, that in general, the BOT-its algorithm is able to obtain an MCAC in less time than IPOG. Notice that IPOG needs to ensure at each point that all the partially-built test cases are consistent with the SUT constraints (see Section~\ref{sec:ipog}).
This implies a query to the SAT solver for each value that it fixes in horizontal extension, another query to the SAT solver for each tuple that has not been covered and an additional query for each of the tests where a tuple can be covered until it is covered.

In contrast, the BOT-its algorithm only checks the first tuple that it fixes in the test and once it finishes building it\footnote{For each of the fixed parameters, BOT-its performs a \emph{limited} query to the SAT solver, which is much more efficient in terms of time} (see Section~\ref{sec:bot}).
In case it is not, it removes the last fixed parameter and checks the test again until the test is consistent.
Although this method can produce as many SAT queries as the IPOG algorithm in the worst case, we observe that this is not true for the average case.
Therefore, the BOT-its algorithm can potentially reduce the number of SAT queries with respect to IPOG.
This comes with the drawback of having a more incomplete heuristic than IPOG, which can increase the test suite size.
Additionally, the BOT-its algorithm is able to eliminate forbidden tuples more efficiently than IPOG (i.e. it does not require one SAT query for each forbidden tuple).

On the other hand, IPOG obtains better MCAC sizes in general.
This is also observed in~\cite{DBLP:conf/cp/AnsoteguiOT21} for the current CT benchmarks.

As a rule of thumb, we recommend using the IPOG algorithm for SUTs with \emph{simple} constraints and low strength $t$ (which are all the available CT benchmarks until this work).
To try to obtain smaller test suites for these instances, metaheuristic approaches such as~\cite{FU2020106288,9402109} can also be useful.
These approaches take an already-built MCAC (for example, one obtained using IPOG) and try to eliminate test cases by swapping values in the test suite using local search techniques.
For the cases where an optimal MCAC is required there exist several algorithms such as MaxSAT MCAC~\cite{ansotegui_incomplete_2022,AnsoteguiIMJ13} or CALOT~\cite{YamadaKACOB15}.
Notice however that obtaining a minimal MCAC for a given SUT and strength $t$ is an NP-Hard~\cite{MaltaisM10} problem, and that only relatively small instances with small strengths (e.g. $t=2$) are suitable for these methods.

In case the generation time is important, for more complex constraints we recommend the BOT-its algorithm and its variants.
Additionally, for higher strengths where the memory consumption of the IPOG algorithm is too high, the PBOT-its variant of BOT-its can mitigate these memory issues.
As a side note, notice also that BOT-its and PBOT-its allow \emph{online} testing of the SUT, as test cases can be applied to the system as soon as they are produced.
In the IPOG algorithm, a test case is not completed until the algorithm finishes.

Trying to find optimal MCACs for these kinds of instances by using MaxSAT MCAC or CALOT can be more challenging.
Unlike in IPOG or BOT-its, these other algorithms need to encode a copy of the SUT constraints for each test, making these approaches impractical when having large SUT constraints (even for $t=2$).

Similarly, metaheuristic approaches can also suffer when dealing with \emph{hard} SUT constraints, as for each change that they perform in the test suite, they must ensure that it is consistent with the SUT constraints by querying the SAT solver.


\section{Conclusions and Future Work}
\label{sec:conclusions}

We have presented a new generator for SUT benchmarks. Thanks to its original and simple design it is possible now to have access, from the CT research community to many different kinds of structures contained in SAT instances that do represent real-world problems. In particular, we can now generate SUT benchmarks of a given size and constraint hardness that will ease the development of new CT tools. Additionally, this approach can be easily adapted to take advantage of the available instances in other constraint programming or operation research formalisms. Finally, an important contribution of this work is to have provided a first detailed study of available CT tools. This study helps to characterize better when we should use a given CT tool and therefore increase the robustness of our testing strategies when facing a new SUT. 

As future work, we will extend SUT-G to generate SUTs with Mixed domains, which can be achieved easily by using CSP instances, MIP instances or even Minizinc instances (formalisms that allow the usage of finite-domain variables) as input to SUT-G instead of SAT instances.
Alternatively, we could detect CSP variables modelled by SAT variables in a SAT instance~\cite{ansotegui_2004}.

A more challenging task is to analyze how the SUT constraints (on the selected SUT parameters) restrict the Covering Array number for a given strength. That is to be able to build SUTs for a given strength whose optimal Covering Array size is guaranteed to be within some boundaries.


\bibliographystyle{unsrtnat}
\bibliography{references}

\end{document}